\documentclass[fleqn,usenatbib]{mnras}

\usepackage{newtxtext,newtxmath}

\usepackage[T1]{fontenc}

\DeclareRobustCommand{\VAN}[3]{#2}
\let\VANthebibliography\thebibliography
\def\thebibliography{\DeclareRobustCommand{\VAN}[3]{##3}\VANthebibliography}

\usepackage{graphicx}	
\usepackage{amsmath}	
\usepackage{tabu}

\newcommand{\add}[1]{\textcolor{black}{#1}}

\title[QSO-host galaxy decomposition]{Spatially resolved dust properties and quasar-galaxy decomposition of \add{a} HyLIRG at \add{$z=4.4$}}

\author[T. Tsukui et al.]{
Takafumi Tsukui,$^{1,2}$\thanks{E-mail: tsukuitk23@gmail.com (TT)}
Emily Wisnioski,$^{1,2}$
Mark R. Krumholz$^{1,2}$
and Andrew Battisti$^{1,2}$
\\
$^{1}$Research School of Astronomy and Astrophysics, Australian National University, Cotter Road, Weston Creek, ACT 2611, Australia\\
$^{2}$ARC Centre of Excellence for All Sky Astrophysics in 3 Dimensions (ASTRO 3D)\\
}

\date{Accepted 2023 May 10. Received 2023 May 8; in original form 2023 February 14}

\pubyear{2023}

\begin{document}
\label{firstpage}
\pagerange{\pageref{firstpage}--\pageref{lastpage}}
\maketitle

\begin{abstract}
We report spatially resolved dust properties of the quasar host galaxy BRI 1335-0417 at redshift $z = 4.4$ constrained by the ALMA observations. The dust temperature map, derived from a greybody fit to rest-frame 90 and 161~$\mu$m continuum images, shows a steep increase towards the centre, reaching $57.1 \pm 0.3$ K \add{and a flat median profile at the outer regions of $\sim$38 K}. Image decomposition analysis reveals the presence of a point source in both dust continuum images \add{spatially coincident with} the highest temperature peak and the optical quasar position, which we attribute to warm dust heated by an active galactic nucleus (AGN). We show that a model including this warm component along with cooler dust heated by star formation describes the global SED better than a single component model, with dust temperatures of 87.1$^{+34.1}_{-18.3}$ K (warm component) and 52.6$^{+10.3}_{-11.0}$ K (cold component). The star formation rate (SFR) estimated from the cold dust component is $1700_{-400}^{+500} M_\odot$ yr$^{-1}$, a factor of three smaller than previous estimates due to a large AGN contribution ($53^{+14}_{-15}$\%). The unresolved warm dust component also explains the steep temperature gradient, as the temperature profile derived after the point source subtraction is flat. The point source subtraction also reduces the estimated central SFR surface density $\Sigma_{\mathrm{SFR}}$ by over a factor of three. With this correction, spatially resolved measurements of $\Sigma_{\mathrm{SFR}}$ and the surface gas mass density $\Sigma_{\mathrm{gas}}$ form a roughly linear sequence in the Kennicutt-Schmidt diagram with a constant gas depletion time of 50-200 Myr. \add{The demonstrated AGN-host galaxy decomposition reveals the importance of spatially resolved data for accurate measurements of quasar host galaxy properties, including dust temperature, star-formation rates, and size.}
\end{abstract}




\section{Introduction}\label{sec:intro}

Star formation activity in the Universe peaked at redshift $2<z<4$ marking a critical stage of galaxy formation \citep{Shapley2011-nx, Madau2014-hn}, during which more massive galaxies are thought to form their stellar mass earlier \citep{Renzini2006-xr}. Hyper-luminous infrared galaxies (HyLIRGs, defined as galaxies with $L_{\mathrm{IR}}>10^{13}~L_{\odot}$) are among the most extreme galaxies found during this epoch, and are thought to host star formation rates (SFR) of over $1000~M_{\odot} \mathrm{yr}^{-1}$. Such objects are suggested to form a majority of stellar mass in the most massive elliptical galaxies in the local Universe over 100~Myr \citep{Narayanan2015-dw}. This extraordinary phase of star formation and subsequent quenching largely decides the fate of early forming massives galaxies and their central black holes \add{(BH)}. In this context, it represents the most important life event of massive galaxy formation. However, we do not yet understand the detailed driving mechanism of this extreme phase of star formation.

One possibility is the gas-rich major merger paradigm, which suggests dynamic evolution of the system with rapid gas infall towards the centre driven by collisions or external gravitational torques of merging galaxies \citep{Sanders1996-gw}. The rapid radial inflow then triggers a nuclear starburst and feeds gas onto the central black hole activating an active galactic nucleus (AGN). The initially dust-obscured AGN removes gas and dust within $\sim 100$ Myr after the peak SFR \citep{Hopkins2012-ru, Davies2007-bd}, at which point the AGN becomes visible as an optical quasar \citep{Hopkins2008-zh, Hopkins2008-fq}. The powerful winds driven by the AGN and the burst of star formation eventually eject the remaining galactic gas and prevent subsequent accretion, leading to the rapid cessation of star formation \citep{Silk1998-gs, Somerville2015-zi}. An alternative to gas-rich major mergers suggested by recent studies is that continuous rapid accretion of gas and satellite galaxies from the large gas reservoir of the cosmic web makes an important contribution to fueling starbursts and the growing black holes \citep{Umehata2019-xg, Mitsuhashi2021-ay, McAlpine2019-oe}. Large gas fractions produced by rapid accretion cause the galactic disk to become gravitationally unstable and form non-axisymmetric substructures (such as clumps, a bar, or spiral arms; \citealt{Hodge2019-uo}), which provide gravitational torques to transfer angular momentum and induce a further inflow of gas \citep[violent disk instability;][]{Dekel2014-oh, Inoue2016-um}.

In either scenario, the formation of the central black hole and the buildup of stellar mass in the surrounding galaxy are co-regulated, with
AGN and star formation feedback 
acting together to produce the observed tight correlation between central black hole masses and host galaxy properties such as central velocity dispersion and bulge stellar luminosity \citep[e.g.,][]{Magorrian1998-iw, Ferrarese2000-pm, Gebhardt2000-xx}. This remarkably tight correlation may encapsulate a fossil record of their co-evolution \citep{Kormendy2013-im}. Early, rapidly evolving galaxies at redshift $z > 2$ are presumably on their way to establishing this BH-host relationship \citep{Izumi2019-de, Pensabene2020-ps}. Studying them, therefore, provides an important probe of the interconnection of AGN and the surrounding star formation.

Our understanding of such starburst galaxies\footnote{\add{Starburst galaxies are commonly referred to galaxies which exhibit the star formation rate three or four times above the tight star formation rate - stellar mass relation which majority of starforming galaxies form \citep[e.g.,][]{Elbaz2011-ec, Rodighiero2011-cz, Schreiber2015-gc}.}} is primarily based on their star formation rate. This quantity has been derived by assuming the measured far-infrared (FIR; $\lambda_{\mathrm{rest}}>100\micron$) emission is mainly due to the cold dust component heated by young massive stars with negligible contribution from the warm dust component heated by AGN (which typically dominates at rest-frame mid-infrared, MIR, $5<\lambda_{\mathrm{rest}}<30\micron$; \citealt{Mullaney2011-sn, Honig2017-sv, Stalevski2016-em}). FIR photometric data are \add{usually} fitted with greybody functions to constrain the dust mass \add{$M_{\mathrm{dust}}$,} temperature \add{$T_{\mathrm{dust}}$}, \add{and the power-law slope at a longer wavelength (opacity index $\beta_{\mathrm{dust}}$)}, which together determine the functional shape. The total infrared (TIR) luminosity $L_{\mathrm{TIR}}$ is then derived by integrating the function over the wavelength range of 8~$\micron$ to 1000~$\micron$, and the star formation rate is derived using an $L_{\mathrm{TIR}}$ to SFR calibration \citep{Kennicutt1998-lm}. If there are photometric data points available over rest-frame MIR to FIR, the standard procedure is to decompose the SED into warm and cold dust components by multi-component SED fitting and to estimate the SFR after removing the contribution from the warm dust presumably heated by AGN \citep[e.g.,][]{Farrah2003-cq, Kirkpatrick2012-sy, Leipski2014-wq, Kokorev2021-yp}. 

However, uncertainty regarding the potential contribution of AGN heating to $L_{\mathrm{TIR}}$ has hindered the understanding the exact nature of high-redshift starburst galaxies. There is growing observational evidence that the AGN contribution to $L_{\mathrm{TIR}}$ is \textit{not} negligible in all galaxies, but instead increases with total galactic luminosity \citep{Alonso-Herrero2012-xq, Nardini2010-rk, Yuan2010-ej, Stanley2017-dc}. Therefore, particularly for the brightest population of HyLIRGs, the nature of the heating source for the thermal dust radiating at FIR wavelengths remains an open question \citep{McKinney2021-zd, Di_Mascia2022-gb} --- does the AGN dominate only for the central region? or does it heat the entire galaxy? \citep{Symeonidis2021-fe}. To date, most attempts to separate these components have relied on spatially unresolved data, for which estimates of the AGN contribution depend on assumptions (e.g., spectral templates) made during SED modelling. Solving the problem requires a thorough investigation of spatially resolved dust properties (e.g., temperature, mass, and optical depth) and spatial separation of the star formation-dominated host galaxy from the AGN-dominated central region. The need for spatially-resolved studies is underscored by the case of the heavily obscured nearby galaxy Arp 220, for which a significant fraction of $L_{\mathrm{TIR}}$ (33\%) originates from a compact region with a dust temperature of $T_{\mathrm{dust}}=200$~K and a radius of 15 pc, suggesting a significant AGN contribution to $L_{\mathrm{TIR}}$ \citep{Scoville2017-wi}.

Spatially resolved measurements such as those for Arp 220 are \add{however} challenging at high redshift. Deriving the temperature distribution as a function of position requires multiple continuum band images near the greybody peak. The most commonly-used instruments for studying the thermal dust properties in the previous decade were the \textit{Herschel \add{Space Observatory (Herschel)}} photometers PACS (from 70 to 160 microns) and SPIRE (from 250 to 500 microns), which provided a spatial resolution (FWHM = 36 arcsec) sufficient only to resolve nearby galaxies \citep{Galametz2012-yy}. Recently, however, the Atacama Large Millimeter/submillimeter Array (ALMA) has started to provide frequency coverage near the peak of the greybody function for $z>4$ galaxies) with enough angular resolution to permit spatially-resolved analysis. While spatially resolved pixel-by-pixel temperature maps for such high redshift galaxies are now attainable in principle, they have not been achieved so far in practice\add{. This is} mainly because high angular resolution observations with a sufficient signal-to-noise ratio in at least two frequency bands\footnote{\add{Strictly speaking, only two bands are not sufficient to constrain the three physical parameters of the greybody function, $M_{\mathrm{dust}}$, temperature $T_{\mathrm{dust}}$, and opacity index $\beta_{\mathrm{dust}}$. However, tight constraints can be obtained by assuming the typical opacity index $\beta_{\mathrm{dust}}$ or choosing two bands near the peak of the spectrum which are sensitive to the dust temperature $T_{\mathrm{dust}}$ and insensitive to the $\beta_{\mathrm{dust}}$.}} are demanding even with ALMA for relatively faint high-redshift galaxies\add{. A few exceptions exist where authors have constrained the temperature gradient within individual galaxies:} \citet{Shao2022-ft} derive 1D radial profiles of the dust temperature, mass, and optical depth for a quasar host galaxy at redshift $z=6$ \add{using} azimuthally averaged profiles at two FIR continuum bands\add{, while \citet{Akins2022-mh} derive a pixel-by-pixel temperature map, which shows evidence for a temperature gradient, in a strongly lensed star-forming galaxy at redshift $z=7$.}


In this paper, we deriv\add{e} spatially-resolved temperature maps for BRI1335-0417, a quasar host galaxy at a redshift of $z = 4.4704$ \add{\citep{Guilloteau1997-xe}}, 1.4 Gyr after the Big Bang. This galaxy is one of the brightest unlensed submillimeter sources known at $z>4$ \citep{Jones2016-cj}, and is classified as a HyLIRG with an extraordinary infrared luminosity of $3.1\times10^{13}L_\odot$\citep{Carilli2002-vo}. It was originally identified as an optical QSO by optical imaging and spectroscopy from the Automatic Plate Measuring survey \citep{APTQSO1991, Storrie-Lombardi1996-rw}. The optical QSO position is RA$=204.514232850\add{\pm0.000000092}$ deg, Dec$=-4.543050970\pm\add{0.000000055}$ deg (ICRS) \citep{2016A&A...595A...1G, 2021A&A...649A...1G}. The galaxy hosts a black hole with a mass suggested to be $10^{9.77} M_\odot$, shining at $\sim 36\%$ of its Eddington luminosity \citep{Shields2006-dy}. The star formation rate of the galaxy was estimated to be \add{$5040\pm1300 M_{\odot}$ yr$^{-1}$} from modelling of the spatially-unresolved spectral energy distribution (SED) with thermal dust and synchrotron emission components \citep{Wagg2014-bh}, a rate high enough to deplete its total molecular gas reservoir of $\sim10^{11}M_{\odot}$ yr$^{-1}$ \add{estimated by CO(2-1) line observation} \citep{Jones2016-cj} in only $\sim$20 Myr. 

Although such short depletion times are commonly attributed to gas-rich major mergers, the morphology and kinematics of the galaxy appear to be inconsistent with this scenario. Spatially resolved [C~\textsc{ii}] and dust continuum observations ($\sim$ 1.3 kpc resolution) show clear evidence for a rotating disk and spiral morphology \citep{tsukui2021-vl} with a further analysis of the gas motion indicating the presence of a compact mass structure in the centre of the galaxy. Indeed, BRI 1335-0417 is the highest-redshift galaxy thus far to show a spiral morphology. The spiral structure is visible in [C~\textsc{ii}] line and dust emission, indicating \add{ongoing star formation}. BRI 1335-0417 is not the only extremely luminous high redshift galaxy to show a surprisingly quiet and well-ordered morphology. Contrary to earlier views that cold disks and spirals only begin to appear at $z<2$ \citep{Elmegreen2014-xx}, ALMA and JWST observations now suggest a surprisingly earlier epoch of galaxy settling at $z\sim 2-4$, with cold gas disks found at $z \sim 4$ \citep[ALMA;][]{Neeleman2020-zu, Rizzo2020-yc, Rizzo2021-ty, Lelli2021-hh, tsukui2021-vl}, stellar spiral structure detected in passive galaxies at $1<z<3$ \citep[JWST;][]{Fudamoto2022-tz}, and grand design barred spirals already in place at $z \sim$2 \citep[JWST;][]{Guo2022-dr}.

This contradiction between the ultra-short depletion time and the morphology of BRI 1335-0417 suggests that it is worth revisiting the high SFR previously estimated from an unresolved SED by using the spatially-resolved data to which we now have access. Doing so may provide insight not just on this particular source, but more broadly on the driving mechanism of high-$z$ starburst\add{s}, early build-up of black holes and stellar bulges, and evolutionary links from the cold gas disks at $z=4$ to stellar spirals at $z \sim 2-3$. With this motivation in mind, this paper aims to (1) derive a resolved dust temperature map and clarify the heating source of the dust, (2) separate the warm dust heated by AGN and cold dust associated with the host galaxy and (3) study the star formation rate distribution under the effects of the central quasar in BRI 1335-0417. We present new Band 9 \add{($\sim484\micron$)} and Band 4 \add{($\sim2080\micron$)} ALMA observations in addition to the earlier Band 7 \add{($\sim869\micron$)} data presented in (\citealt{tsukui2021-vl}; PI=Gonz\'alez L\'opez, Jorge). Multiple continuum images provide a number of resolution elements over the galaxy, making it possible to investigate the spatially resolved physical properties of dust (such as temperature and optical depth.) The paper is organized as follows. In Section \ref{sec:obs}, we describe the observation\add{s} and the data reduction. In Section \ref{sec:result}, we present the results of dust SED modelling applied to each pixel of the spatially resolved continuum images. In Section \ref{sec:result2}, we describe our image decomposition method (point source and host galaxy) followed by the panchromatic SED modelling of BRI 1335-0417 with the help of the decomposition results. In Section \ref{sec:discussion}, we present the discussion of the result. In Section \ref{sec:conclusion}, we summarize and conclude the paper. 

Throughout the paper, we adopt a flat lambda cold dark matter ($\Lambda$CDM) cosmology with a present-day Hubble constant $H_0=70$ km s$^{-1}$ Mpc$^{-1}$, and a density parameter of pressureless matter $\Omega_M=0.3$, providing an angular size distance $D_\mathrm{A}=1375$ Mpc and a luminosity distance $D_\mathrm{L}=40202$ Mpc at the redshift of 4.4074.

\section{Observations and Data reduction} \label{sec:obs}
\add{\subsection{ALMA imaging}}
ALMA Band 4, Band 7, and Band 9 observations of BRI 1335-0417 were carried out as part of the programs \#2017.1.00394.S, and \#2018.1.01103.S (PI=González López, Jorge). 
These observations targeted emission lines, including CO(7-6) (Band 4) and [C~\textsc{ii}] (Band 7), along with underlying continuum emission at observing wavelength 2080~$\micron$ in Band 4, 869~$\micron$ in Band 7, and 484~$\micron$ in Band 9, corresponding to rest-frame 385~$\micron$, 161~$\micron$, 90~$\micron$, respectively. We performed standard calibration and data reduction using the Common Astronomy Software Application (\textsc{casa}; \citealt{CASA_Team2022-cl}) pipeline\footnote{We used the same \textsc{casa} version as used in the quality assurance at the ALMA Observatory, which is 5.1.1-5 for Band 7, and 5.4.0-70 for Band 4 and Band 9 data}. The flux and bandpass were calibrated using the quasars J1337-1257 for Band 4 and Band 7, and J1256-0547 for Band 9 data. The phase was calibrated using the quasars J1332-0509 for Band 4 and J1336-0829 for Band 7 and Band 9. We identified line-free channels using the \textsc{hif}\_\textsc{findcont} task in \textsc{casa} and additionally removed channels affected by atmospheric absorption. We estimated the flux density of the underlying continuum emission by fitting a linear function to the identified line-free channels and subtracted it from the data cube in the visibility plane. 
We then imaged the line-free channels to produce continuum images and the continuum-subtracted data to produce emission line cubes. The visibility data were weighted using a Briggs weighting scheme with a robust parameter of 1.0 for the CO(7-6) line data cube to improve the sensitivity and 0.5 for others, which provides a good compromise of spatial resolution and sensitivity \citep{Briggs1995-yt}. The resulting images and cube information such as angular resolution and point source sensitivity are summarized in Table~\ref{tab:tab1}. The emission line flux maps are made by summing the velocity channels from $-400$ to 400km s$^{-1}$. The mean velocity and the velocity dispersion of the emission line are extracted by fitting a single Gaussian to the spectrum at each pixel. We only use pixels where the signal is detected more than 4 $\sigma$ over at least 4 channels.

In this paper, the same phase centre and pixel size are used to image the visibility data so that all data products have the same pixel coordinates, with absolute positional accuracy of 10 to 20 mas.\footnote{ALMA Technical Handbook \url{https://almascience.nrao.edu/documents-and-tools/cycle9/alma-technical-handbook}} Before deriving the physical parameters using two or more images (e.g., dust SED modelling), we convolved images to have the same spatial resolution with the smallest common beam size, using \textsc{common}\_\textsc{beam} in the \textsc{spectral-cube} package. 
We similarly convolve all other data products to the same resolution, so that physical parameters derived (e.g., velocity dispersions derived from emission line cubes) are resolution-matched. 
The matched resolution of the [C\textsc{ii}] cube, rest-frame 161~$\micron$, and 90~$\micron$ continuum images is 0.21"$\times$0.18" at P.A.=114 deg (beam area of 1.92 kpc$^2$, and effective radius of 0.78kpc). \add{Analysis and visualisation of data in this paper are done at this resolution except for the radial profile analysis 
carried out in Section~\ref{ssec:mass_profiles}, which includes the poorer resolution CO(7-6) data. For the comparison with CO(7-6) all data are convolved to the common} resolution of 0.24"$\times$0.18" at P.A.=94 deg (beam area of 2.3 kpc$^2$ and effective radius of 0.85kpc). \add{The effects of the beam on our spatially resolved SED analysis are also discussed in Section \ref{subsec37}.}

\begin{table*}
    \centering
    \begin{tabular}{c c c c c c}
         \hline\hline
          Data & Rms noise & Synthesized beam & Subtended area per beam & Rms noise in SFR \\
           & ($\mu$Jy beam$^{-1}$) & maj(")/min(")/P.A.(deg) & kpc$^2$ beam$^{-1}$ & $M_\odot$ yr$^{-1}$ kpc$^{-2}$\\
          \hline\hline 
          $I_{\lambda_{\mathrm{rest}}=385\mu\mathrm{m}}$ (Band 4)& 6.58 & 0.21/0.14/85 & 1.52 & 11.8\\
          $I_{\lambda_{\mathrm{rest}}=161\mu\mathrm{m}}$ (Band 7)& 25.9 & 0.19/0.16/75 & 1.56 & 7.5\\
          $I_{\lambda_{\mathrm{rest}}=90\mu\mathrm{m}}$ (Band 9) & 186 & 0.21/0.16/-53 &  1.69 & 5.4\\
          \hline
          $\left[\mathrm{C}~\textsc{ii}\right]$ &  389 & 0.19/0.16/82 & 1.59 & -\\
          CO(7-6) &  101 & 0.23/0.16/84 & 1.93 & -\\
         \hline\hline

    \end{tabular}
    \caption{ALMA observation data summary. The root mean square (rms) noise level is measured using the emission-free region of the images. Synthesized beams are reported as (major axis, minor axis, and position angle). Rms noise in the unit of star formation rate is calculated by assuming  optically thin greybody emission with a dust temperature of 39K, a dust emissivity index $\beta=2.14$ (see Sect. \ref{subsec:result2}), and a \citet{Chabrier2003-el} initial mass function.}
    \label{tab:tab1}
\end{table*}

\subsection{\add{HST imaging}}
\begin{figure}
\centering
\includegraphics[width=\columnwidth]{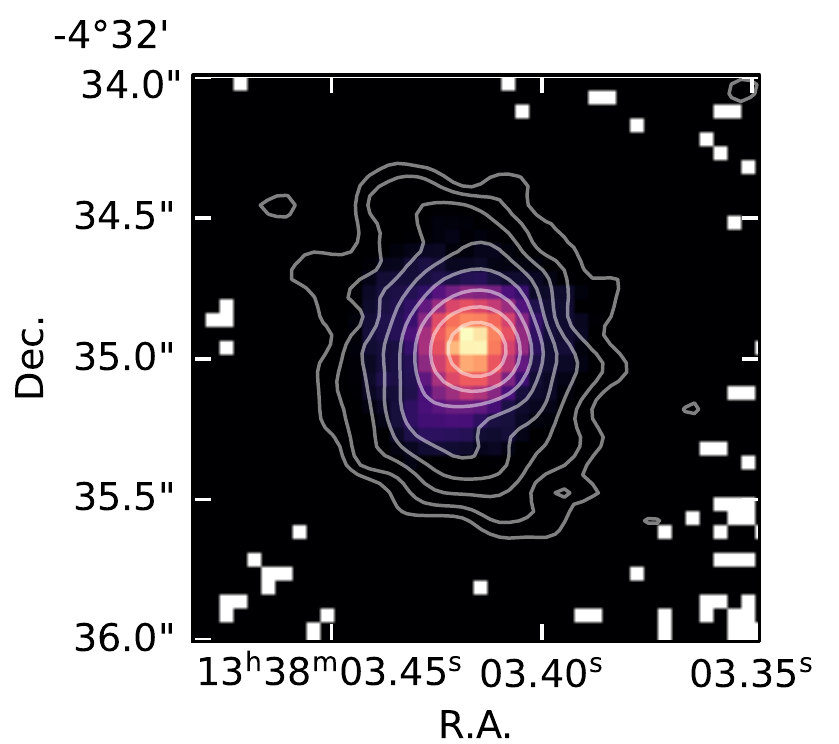}
\caption{\add{\textit{HST} STIS/50CCD image overlain with the ALMA Band 7 image contour. The optical band emission is consistent with the point source. In contrast, the ALMA Band 7 image provides the extended morphology of the host galaxy.\label{fig:fig0}}}
\end{figure}

\add{We retrieved the STIS/50CCD image of BRI 1335-0417 from the \textit{Hubble Space Telescope} (\textit{HST}) archive. The data were taken as a part of program GO-8572 (PI=L.~Storrie-Lombardi) in January 2001, consisting of the four dither exposures with each integration time of 645s. The detector has broad sensitivity from 2000 to 10300\AA. Therefore, we only use the data for visualization purposes to assist the interpretation of the other optical photometric data of the galaxy shown in Table~\ref{tab:taba1}. We processed the fully calibrated sub-exposure images from the \textit{HST} archive applying the geometrical distortion and world coordinate system (WCS) corrections, and then drizzling onto the final pixel grid with the pixel size of 0.05". The target acquisition is based on the Guide Star Catalog GSC 1.0, which is expected to have a pointing accuracy of 1-2" in worst cases. We calibrate the pointing offset of the image by translation to match the image peak of the BRI 1335-0417 to the corresponding Gaia coordinate. We identify the BRI 1335-0417 in the image using three Gaia sources available in the field of view, including BRI 1335-0417. The calibrated offset length of 0.8" is within the expected pointing accuracy. Figure~\ref{fig:fig0} shows the 50CCD image overlaid with the contours of the ALMA Band 7 continuum image. The optical image is consistent with the point source emission from the AGN (see Fig.~\ref{fig:figA0} for a comparison of the radial distribution of the emission and the point spread function). In contrast, the ALMA Band 7 image clearly resolves the extended host galaxy.}

\begin{figure*}
\centering
\includegraphics[width=\textwidth]{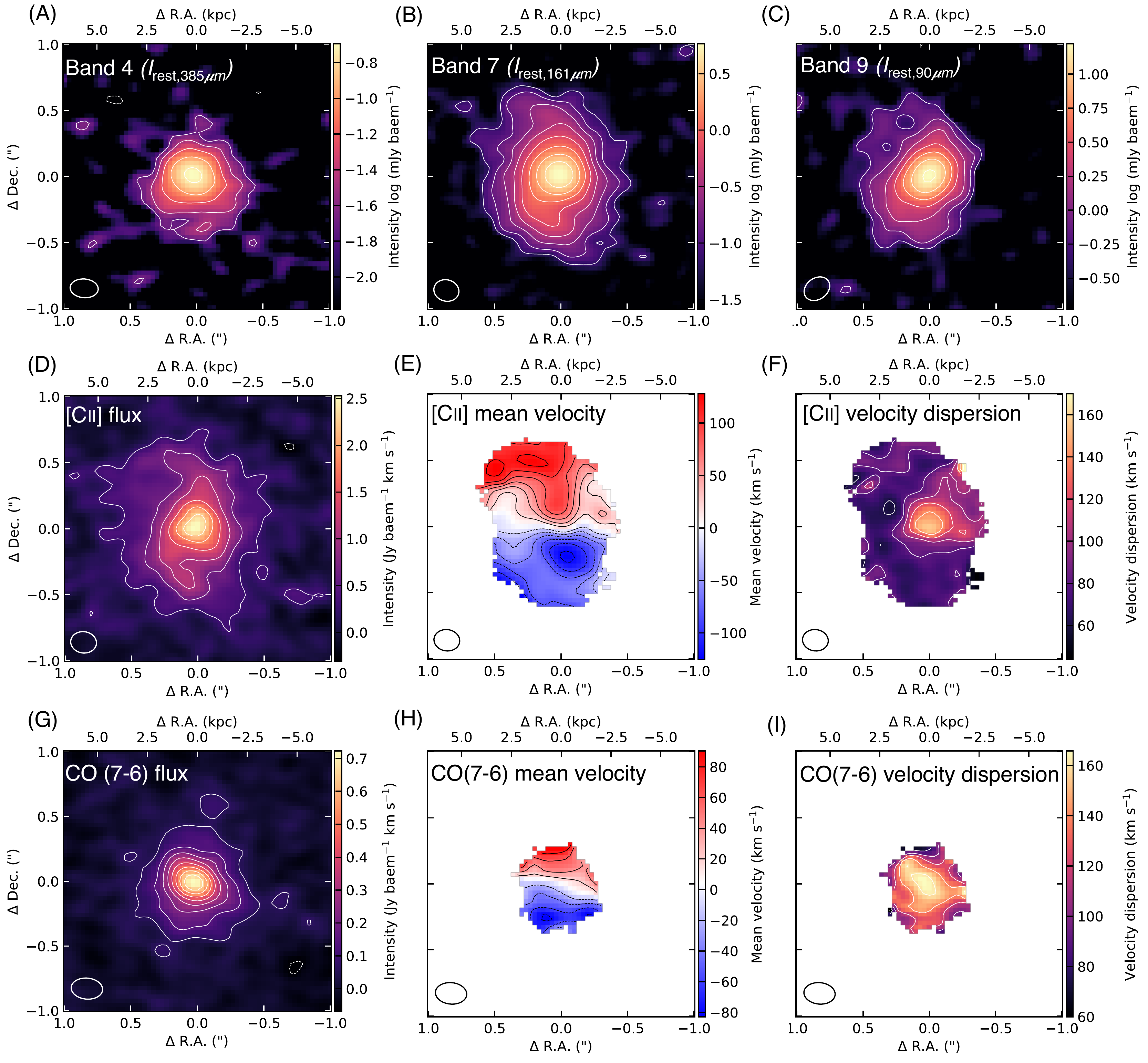}
\caption{ALMA observation for QSO source BRI 1335-0417 at redshift z$\sim$4.4. The top panels show the continuum images at rest-frame  385~$\micron$ (Band 4; A), 161~$\micron$ (Band 7; B), and 90~$\micron$ (Band 9; C) from left to right. The middle panels show the [C\textsc{ii}] integrated flux (D), mean velocity (E), and velocity dispersion maps (F). The bottom panels show the CO(7-6) flux (G), mean velocity (H), and velocity dispersion (I) maps. The mean velocity and velocity dispersion at each pixel are extracted by fitting a single Gaussian to the spectrum. We only use pixels where the signal is detected more than 4$\sigma$ over at least 4 channels. The solid line contours in \add{(}A\add{)}\add{, (B) and }\add{(}C\add{)} are plotted for 3$\sigma\times\sqrt{3}^{n};n=\{0,1,2,3,...\}$ to 27$\sigma$, $140.3\sigma$, and $81\sigma$, respectively. The solid line contours in \add{(}D\add{)} and \add{(}G\add{)} are plotted for $3\sigma+4n; n=\{0, 1, 2, 3, ...\}$ to 23\add{$\sigma$} and 31\add{$\sigma$} respectively. Dotted line contours in all flux maps show $-3\sigma$. Contours in velocity and velocity dispersion map are shown every 20 km s$^{-1}$, where plus and minus values are shown in solid and dotted lines, respectively. The size of the synthesized beam (FWHM) is shown in the lower-left corner of each map. \label{fig:fig1}}
\end{figure*}

\section{Spatially resolved dust properties} \label{sec:result}
\subsection{The resolved FIR continuum and [C~\textsc{ii}] emission}\label{subsec:result1}
In Figure~\ref{fig:fig1}, we show continuum images of rest-frame 385~$\micron$, 161~$\micron$, 90~$\micron$, which we denote $I_{\lambda_{\mathrm{rest}}=385\mu\mathrm{m}}$, $I_{\lambda_{\mathrm{rest}}=161\mu\mathrm{m}}$, $I_{\lambda_{\mathrm{rest}}=90\mu\mathrm{m}}$, respectively in the paper, together with the total flux and mean velocity and velocity dispersion derived from the [C~\textsc{ii}] and CO(7-6) line data. In all images of this paper, the RA and Dec offsets are given relative to the position RA$=204.514232$ deg and Dec$=-4.543055$ deg, which coincides with the peak of the continuum images and the quasar position within the positional accuracy of ALMA. The continuum emission is more centrally concentrated than the [C~\textsc{ii}] line emission, as the 385~$\micron$, 161~$\micron$, 90~$\micron$ continuum images and their contours are shown in log scale while the [C~\textsc{ii}] image is shown in linear scale (We also show linear-scaled continuum images below, in Section \ref{sec:result2}). This difference is commonly seen in other spatially resolved observations of quasar host galaxies \citep[e.g.,][]{Walter2022-uh, Shao2022-ft}. As presented in \citet{tsukui2021-vl}, both $I_{\lambda_{\mathrm{rest}}=161\mu\mathrm{m}}$ and [C~\textsc{ii}] images show Z-shaped spiral structures. The $I_{\lambda_{\mathrm{rest}}=90\mu\mathrm{m}}$ image also has a disk-like morphology with a major axis similar to that in $I_{\lambda_{\mathrm{rest}}=161\mu\mathrm{m}}$, but a fainter spiral arm feature. The $I_{\lambda_{\mathrm{rest}}=385\mu\mathrm{m}}$ \add{is less sensitive to SFR which leads to lower S/N}, so the emission is not detected in the outer part of the galaxy. The mean [C~\textsc{ii}] velocity map is qualitatively consistent with the moment 1 map (intensity-weighted velocity of the spectral line) presented in \citet{tsukui2021-vl}, but provides more accurate measurements than the moment 1 map, which can easily be affected by noise in the emission-free channels. The [C~\textsc{ii}] velocity dispersion map shows a triangle-shaped velocity-enhanced region extending from the centre to the North-West direction. Such an asymmetric structure cannot be explained by the velocity gradient of a regularly rotating motion, suggesting that the origin may not be gravitational. 

\subsection{Dust SED modelling}\label{subsec:result2}
To derive dust properties from our observed spatially resolved FIR fluxes, we use a greybody function to describe the flux. 
For a uniform region of dust with an equilibrium temperature $T_\mathrm{dust}$ at redshift $z$, solving the equation of radiative transfer yields a predicted CMB-subtracted flux within the ALMA beam \citep{Walter2022-uh}
\begin{equation}\label{eq:eq1}
    F_\nu=\Omega_a\times[B_\nu(T_{\mathrm{dust,z}})-B_\nu(T_{\mathrm{CMB,z}})]\times[1-\exp(-\tau_\nu)](1+z)^{-3},
\end{equation}
where $\Omega_a$ is the solid angle of the ALMA synthesized beam in steradians, $B_\nu$ is the Planck function, $T_{\mathrm{dust,z}}$ is the dust temperature, and $T_{\mathrm{CMB},z}=2.73(1+z)$ K is the CMB  temperature at \add{a given} redshift\add{,} $z = 4.4704$ \add{in our case}. $\tau_\nu$ is optical depth, which is related to the dust mass $M_{\mathrm{dust}}$ by $\tau_\nu=\kappa_0(\nu/\nu_0)^\beta M_{\mathrm{dust}}A^{-1}$, where $A=\Omega_a D_\mathrm{A}^2$ is the physical area subtended by the beam, $\beta$ is the dust opacity index, and $\kappa_0=5.1$ cm$^2$ g$^{-1}$ is the dust opacity at a reference frequency $\nu_0=1199.1$ GHz \citep{Draine2007-ou}. Note that while the observed flux depends on the actual dust temperature at the redshift of the source, $T_{\mathrm{dust},z}$, for convenience we compute and report the intrinsic dust temperature corrected for the CMB heating, $T_{\mathrm{dust}}$, which is the temperature the dust would have at redshift zero, where CMB heating is much smaller. \add{As described in \citet{Da_Cunha2013-yv}, t}hese two temperatures are related \add{as},
\begin{equation}\label{eq:eq2}
T_{\mathrm{dust}, z}=\left\{T_{\mathrm{dust}}^{\beta+4}+T_{\mathrm{CMB}, z=0}^{\beta+4}\left[(1+z)^{\beta+4}-1\right]\right\}^{\frac{1}{4+\beta}}.   
\end{equation}

\subsection{Global SED of the galaxy}
\begin{figure*}
\centering
\includegraphics[width=0.6\textwidth]{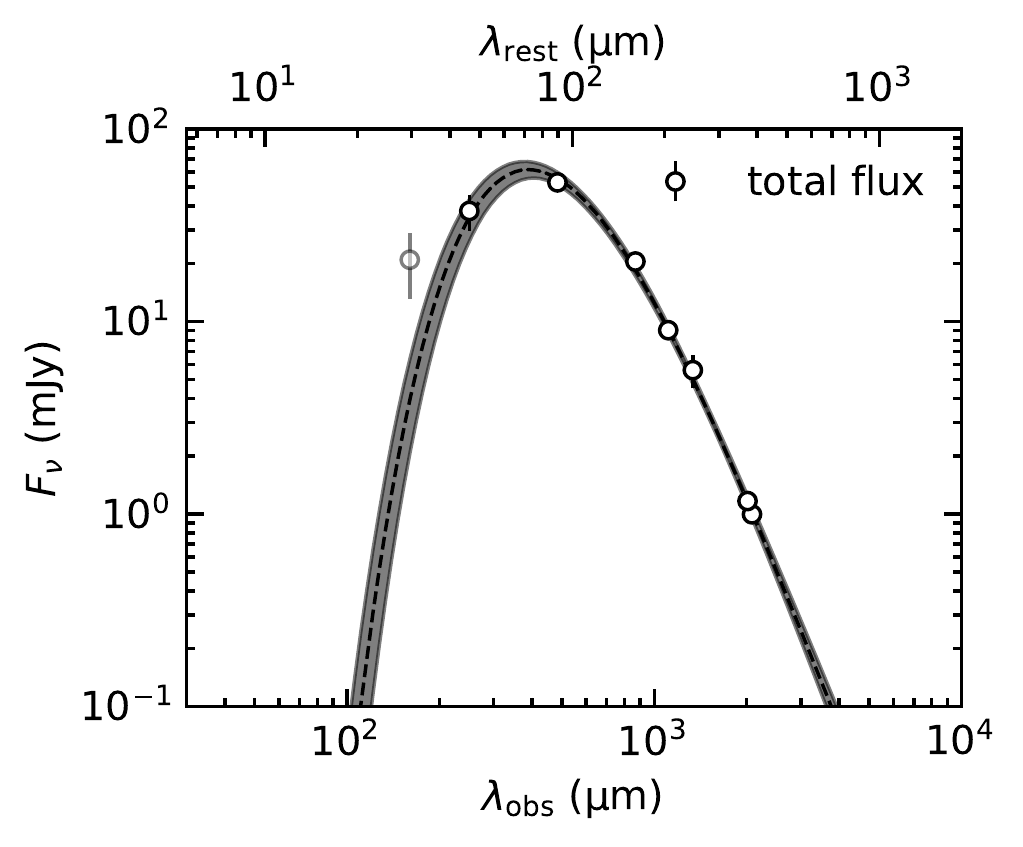}
\caption{The SED of BRI 1335-0417 fitted with a single greybody in the optically thin limit, with the best-fit parameters of $T_{\mathrm{dust}} = 39.0_{-2.9}^{+3.4}$K and $\beta=2.14\pm0.17$. The available data points are listed in Table~\ref{tab:taba1}, where the greybody fit is performed on bands longer than 160~$\micron$ (rest-frame 36~$\micron$); we exclude the 160~$\micron$ data \add{(the grey point)} from the fit because a single greybody fit cannot reproduce it. The TIR luminosity derived from the fit is $L_{\mathrm{TIR}}=2.7_{-0.3}^{+0.4}\times 10^{13} L_\odot$, consistent with the previously reported value for this galaxy of 3.1$\times10^{13}L_\odot$ \citep{Carilli2002-vo}. \label{fig:fig2}}
\end{figure*}

In Fig.~\ref{fig:fig2} we show the total (spatially integrated over the galaxy) SED of BRI 1335-0417 with our new flux measurements at rest-frame 385~$\micron$, 161~$\micron$, 90~$\micron$. We measured the integrated \add{flux densities} by summing over the 2"$\times$2" region shown in Fig.\ref{fig:fig1}. The available data points are listed in Table~\ref{tab:taba1}. We excluded available photometric measurements from the \textit{Herschel} SPIRE 350~$\micron$ and 500~$\micron$ bands because BRI 1335-0417 is not sufficiently separated from a nearby bright source at \textit{Herschel}'s angular resolution (see each photometric image in Fig.~\ref{fig:figa1}). To constrain the global dust properties of the galaxy, we fit the SED with a single greybody in the optically thin limit ($\tau_\nu \rightarrow 0$, large $\Omega_a$). In this limit, Eq.~\ref{eq:eq1} becomes,
\begin{equation}\label{eq:eq3}
     F_{\nu}=(1+z)D_L^{-2} \kappa_{\nu}M_{\mathrm{dust}}[B_{\nu}\left(T_{\mathrm{dust,z}}\right)-B_{\nu}\left(T_{\mathrm{CMB,z}}\right)].
\end{equation}
In the fitting, we treat $T_{\mathrm{dust}}$, $M_{\mathrm{dust}}$, and $\beta_{\mathrm{dust}}$ as free parameters. We find that a single greybody fit cannot reproduce the flux observed in the \textit{Herschel} 160~$\micron$ band (rest-frame 36~$\micron$), where warm dust is expected to dominate. Therefore, we repeat the fit with the 160~$\micron$ band flux excluded; doing so yields a total dust mass of $1.9^{+0.3}_{-0.3}\times10^9M_{\odot}$, a dust temperature $T_{\mathrm{dust}} = 39.0_{-2.9}^{+3.4}$K and an emissivity index $\beta=2.14\pm0.17$, which describes the power law slope of the greybody function in the Rayleigh-Jeans tail ($\propto \lambda^{-2-\beta}$ at $\lambda\gg100\micron$). The derived emissivity index and the computed total FIR luminosity $L_{\mathrm{TIR}}=2.7_{-0.3}^{+0.4}\times 10^{13} L_\odot$ are consistent with the previously derived values for this galaxy, $\beta=1.89\pm0.23$ \citep{Wagg2014-bh} and $L_{\mathrm{TIR}}\sim3.1\times 10^{13} L_\odot$ \citep{Carilli2002-vo}. The gas-to-dust mass ratio is found to be 54.2$\pm 9.3$ if we adopt a \add{total} gas mass derived by \citealt{Jones2016-cj} from CO (2-1) using the standard conversion factor for SMGs $\alpha_{\mathrm{CO}}=0.8$ M$_\odot$ pc$^{-2}$/ (K km s$^{-1}$) \add{\citep{Bolatto2013-ch}} and a ratio of $\mathrm{r}_{21}=L_{\mathrm{CO}(2 \rightarrow 1)}^{\prime} / L_{\mathrm{CO}(1 \rightarrow 0)}^{\prime}=0.85$ for SMGs \citep{Carilli2013-im}. 

\add{The flux density excess, which cannot be captured by a single greybody function in from rest-frame $\sim8$ to $50\micron$, has previously been attributed to the presence of higher temperature components particularly in nuclear regions \citep{Casey2014-pu}. Different prescriptions have been employed in the literature to reproduce the excess flux, including the sum of two greybody functions with different dust temperatures \citep{Dunne2001-rz, Farrah2003-cq} or a greybody function with the Wien part replaced by the power law function \citep{Casey2012-yt}. In the former method, the warmer component has been commonly attributed to the AGN-heated dust \citep{Farrah2003-cq, Kirkpatrick2012-sy, Leipski2014-wq, Kokorev2021-yp}, an interpretation we will explore in detail with the spatially resolved data in this study. The latter approach of adding a power law component is somewhat phenomenological, but has the advantage that it introduces only a single additional free parameter -- an advantage that is not negligible when the set of measurements available to fit the data is very small, as it is here.}


\subsection{Intrinsic dust properties derived from the spatially resolved dust SED}
Spatially resolved images at different frequency bands allow us to derive the intrinsic dust properties at each individual pixel. The galaxy is spatially resolved with 14, 32, and 20 resolution elements detected at $\geq 3$ $\sigma$ significance in the rest-frame 385~$\micron$ (Band 4), 161~$\micron$ (Band 7), and 90~$\micron$ (Band 9) continuum images, respectively. The rest-frame 161~$\micron$ and 90~$\micron$ bands are close to the peak of the greybody, and their ratio is therefore sensitive to the dust temperature. The spatially resolved rest-frame 385~$\micron$ may help to constrain the emissivity index $\beta$ or the SED slope at a longer wavelength. However, the rest-frame 385~$\micron$ data has the worst sensitivity in terms of the minimal detectable SFR (see Table~\ref{tab:tab1}), and the $3\sigma$ detected emission area is smaller than in the other bands. Therefore, we choose to use the best estimate of $\beta=2.14\pm0.17$ obtained with the global SED modelling as a fiducial value in this paper, and not to include the rest-frame 385~$\micron$ band for spatially resolved SED modelling. With flux densities measured in two bands and the fiducial dust emissivity $\beta=2.14\pm0.17$, we can fully constrain the two remaining free parameters $\Sigma_\mathrm{dust}$ and $T_\mathrm{dust}$ in equations \ref{eq:eq1} and \ref{eq:eq2}. We propagate the uncertainty on the spectral index, $\beta=2.14\pm0.17$, to the derived dust properties via Monte Carlo. 

\begin{figure*}
\centering
\includegraphics[width=\textwidth]{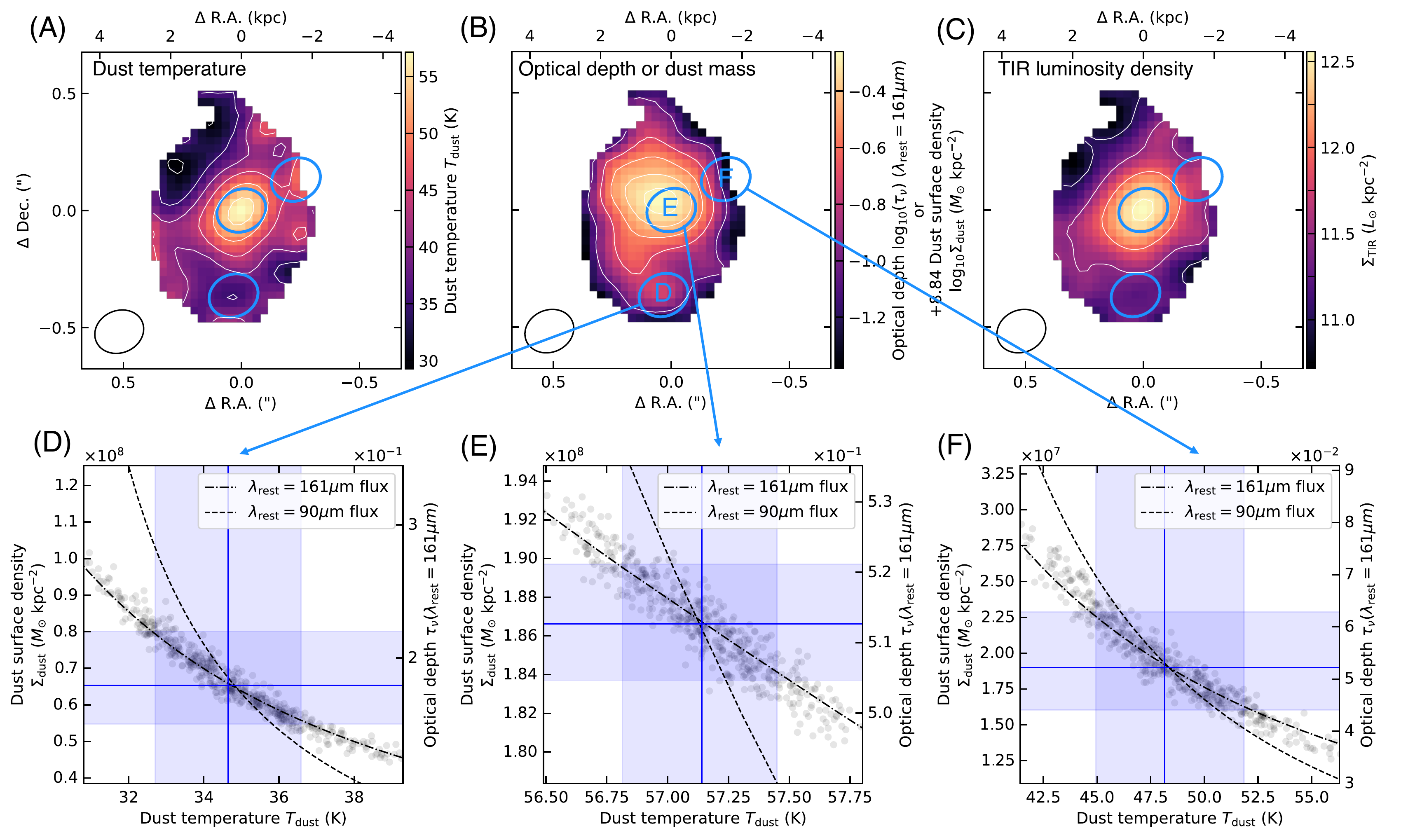}
\caption{The upper panels show the spatial distribution of the best-fit dust temperature (A), optical depth at $\lambda_{\mathrm{rest}}=~161\mu$m or dust mass surface density (B), and surface density of the integrated total IR luminosity over 8-1000~$\mu$m, $\Sigma_{\mathrm{TIR}}$($L_\odot$ kpc$^{-2}$) (C). The lower panels show the isoflux lines of the measured flux density ($\lambda_{\mathrm{rest}}=161~\mu$m ; dot-dashed line, $\lambda_{\mathrm{rest}}=90~\mu$m; dashed line) in the parameter space $\Sigma_{\mathrm{dust}}$ and $T_{\mathrm{dust}}$, for three example regions: (D) the region with the lowest temperature, (E) the central region with the highest temperature, and (F) the region with the second highest temperature. The regions (D-F) are indicated with blue ellipses in panels (A-C) and labels in panel (B). The intersection of the two black lines corresponds to the most likely solution (blue solid lines), while the blue shades enclose the 68th percentile of the resampled distribution (grey points) reflecting the noise in the data and the confidence interval for the fiducial dust emissivity $\beta=2.14\pm{0.17}$ derived in the global SED fitting (Fig.~\ref{fig:fig2}). Contours in (A) are shown every 5K from 30K to 55K. Contours in (B) are shown every 0.1 from 0.1 to 0.5. Contours in (C) are shown for $10^{11}$, $10^{11.5}$, $10^{12}$, and $10^{12.5}$$L_{\odot}$kpc$^{-2}$. The size of the synthesized beam (FWHM) is shown in the lower-left corner of (A-C).\label{fig:fig3}}
\end{figure*}

Figures~\ref{fig:fig3}~(A-C) show the best-fit dust temperature, the optical depth of the dust at rest-frame $161\micron$ (or the dust surface density $\Sigma_\mathrm{dust}$), and the integrated dust luminosity density maps of the galaxy, respectively. We derive our 1 $\sigma$ confidence intervals on the fit parameters ($T_\mathrm{dust}$ and $\Sigma_\mathrm{dust}$) by Monte Carlo resampling: we repeat the fit 300 times using fluxes randomly drawn from a Gaussian distribution centred on the best fit measured value with a dispersion equal to the 1$\sigma$ observational uncertainty, and a spectral index $\beta$ drawn from a Gaussian centred at $\beta = 2.14$ with a dispersion of $0.17$. We report the width of the central 68\% of these 300 trials as our $1\sigma$ uncertainties on $T_\mathrm{dust}$ and $\Sigma_\mathrm{dust}$. We only show and use pixels where the 68\% confidence interval for the dust temperature is smaller than 10K. Figures~\ref{fig:fig3}~(D-F) show how the measured flux densities at rest-frame $161\micron$ (black dot-dashed line) and $90\micron$ (black dashed line) band constrain dust physical parameters $\Sigma_\mathrm{dust}$ and $T_\mathrm{dust}$ at fixed $\beta=2.14$, for three example regions: (D; left) the region with the lowest temperature, (E; middle) the central region that has the highest temperature, and (F; right) the region with the 2nd highest temperature. The intersection of the two black lines corresponds to the most likely solution of $\Sigma_\mathrm{dust}$ and $T_\mathrm{dust}$ and the resampled distributions are shown in grey points. The dust temperature $T_\mathrm{dust}$ and surface density $\Sigma_\mathrm{dust}$ are correlated but still constrained around the intersection of the isoflux lines of the two bands.

The temperature map reveals a central high-temperature region surrounded by a low-temperature region. The highest temperature peak coincides with the peak position of both dust continuum emission images $I_{\lambda_{\mathrm{rest}}=90\mu\mathrm{m}}$ and $I_{\lambda_{\mathrm{rest}}=161\mu\mathrm{m}}$. The dust temperature of the central pixel is well constrained to be $57.1 \pm 0.3$ K, which is higher than typical dust temperatures of quasar host galaxies \citep[$\sim$47K;][]{Beelen2006-fc}, indicating that the central high-temperature peak is likely due to the dust heated by the AGN. The temperature map also shows several high-temperature regions in the outer part of the galaxy. The surface density or optical depth maps show that the central region is moderately optically thick at rest-frame $161~\mu$m and optically thick at rest-frame $90~\mu$m ($\tau>1$). The surface density map shows the southern spiral structure, which corresponds to the relatively low temperature $<$ 40K region in the temperature map.

\begin{figure}
\includegraphics[width=\columnwidth]{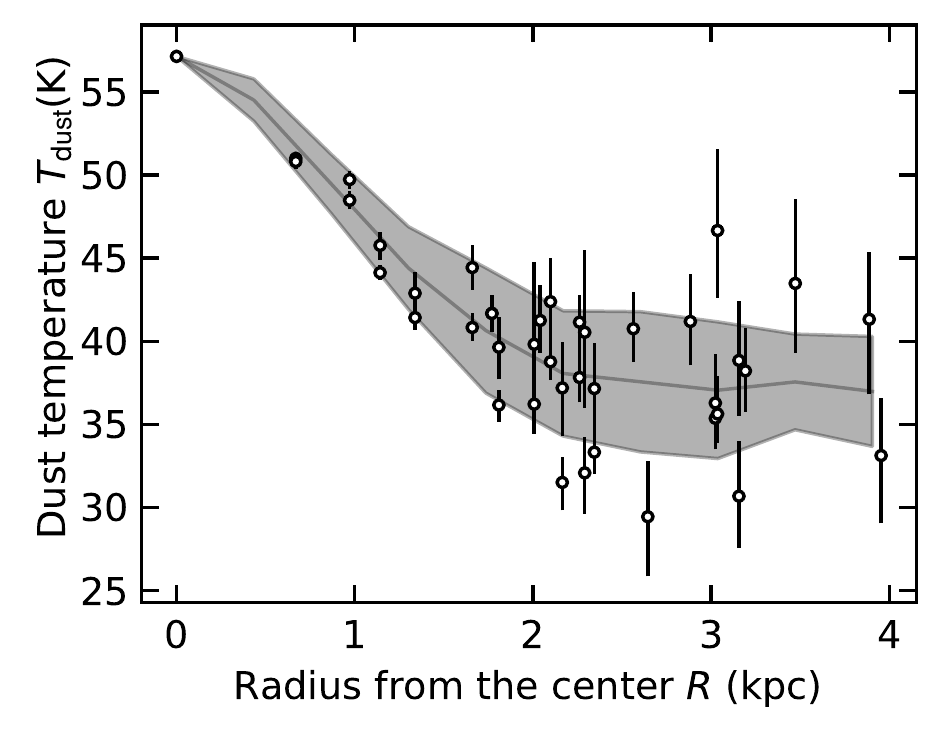}
\caption{Radial profile of the temperature of BRI 1335-0417. The black points with error bars show temperature values, 3 data points per beam extracted from the temperature map (Fig.~\ref{fig:fig3}A) with a regular grid aligned with the central pixel. The grey-shaded region encloses 68\% of all available pixels in radial bins with a width of 1/3 of the beam FWHM. The grey line indicates the median in each bin. The horizontal axis shows the distance from the galactic centre (dust peak position) assuming the disk geometry presented in \citet{tsukui2021-vl}.\label{fig:fig4}}
\end{figure}

Figure~\ref{fig:fig4} shows the radial distribution of the temperature, computed from data points extracted from the image (Fig.~\ref{fig:fig3}A). We calculate the de-projected radius using the thin disk geometry proposed in \citet{tsukui2021-vl} (Position angle 7.6 deg, inclination 37.3 deg), derived from dynamical modelling of [C~\textsc{ii}] emission kinematics with prior constraints from the axial ratio of the continuum image and the kinematic position angle of [C~\textsc{ii}] emission. The derived dust temperature steeply decreases as a function of the galactic radius from the centre to 2.5 kpc. At the outer disk ($\sim3$ kpc), both high- and low-temperature regions can be seen, and the temperature difference is statistically significant, corresponding to the clumpy high-temperature regions seen in the map. At a larger radius, the temperature profile becomes flat with a median temperature of 38K and a large spread of values from 30 to 45 K, roughly consistent with the typical dust temperature of the high redshift starburst galaxies \citep{Magnelli2012-rf} at $L_{\mathrm{TIR}}\sim 10^{13}L_{\odot}$. The presence of the high-temperature regions at a larger radius may be common in other quasar host galaxies and is consistent with the observational signature of J0305-3150 at $z = 6.6$, which shows an increase in the surface brightness ratio $I_{\lambda_{\mathrm{rest}}=175~\mu\mathrm{m}}/I_{\lambda_{\mathrm{rest}}=459~\mu\mathrm{m}}$ with radius \citep{Li2022-tk}, for which the authors suggested two possibilities: radial increase of the temperature or decrease of the opacity. We discuss the possible origin of the outer-disk high-temperature regions in Sec.~\ref{subsec:subsec35}. 

\add{\subsection{Comparison of the derived dust mass profile and other gas mass tracer profiles}\label{ssec:mass_profiles}}

\begin{figure}
\includegraphics[width=\columnwidth]{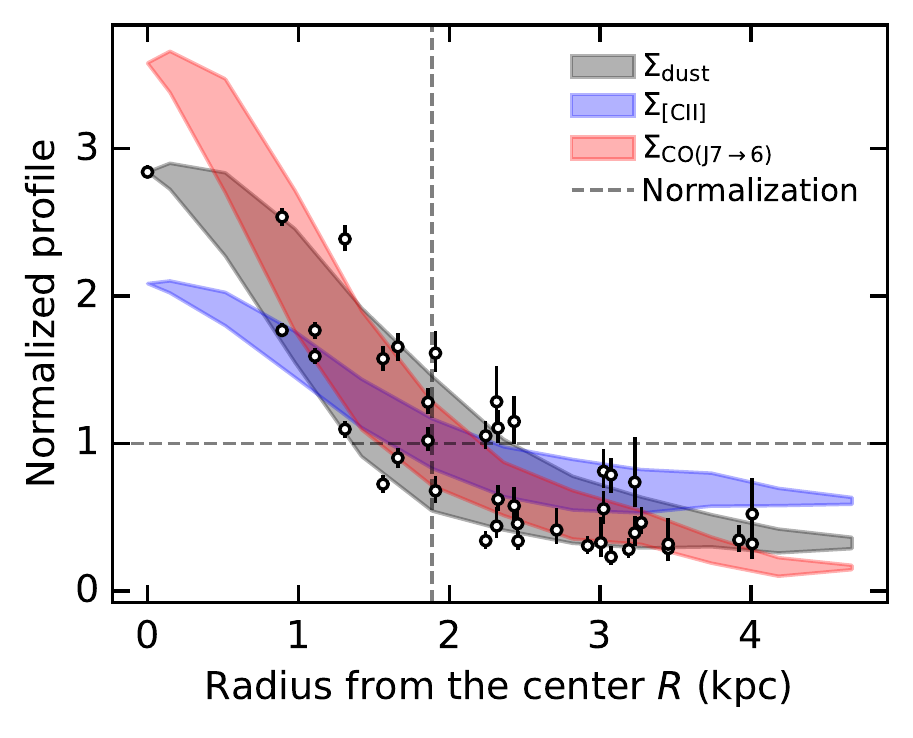}
\caption{Radial profile of normalized gas mass surface densities of BRI 1335-0417. The black points with error bars and black shade show gas surface density assuming the constant gas-to-dust ratio, extracted in the same way as Fig.~\ref{fig:fig4} from the dust surface density map (Fig.~\ref{fig:fig3}B). We also show two additional gas mass estimates from [C~\textsc{ii}] (blue shade) and CO(J=7$\rightarrow$6) (red shade) using a constant mass-to-light ratio. All the profiles are normalized to 1 at a radius of $\sim$2kpc. \label{fig:fig5}}
\end{figure}

Figure~\ref{fig:fig5} shows surface gas mass densities derived by three possible gas mass tracers: emission lines of [C~\textsc{ii}] and CO(7-6), and the SED-derived dust mass. For the two former methods, we assume a constant emissivity per unit mass, while for the latter we assume a constant dust-to-gas ratio, and we normalize the profiles derived from each method to be equal at $\sim2$kpc. Both the dust-derived and CO-derived gas mass profiles are steeper than that estimated from the [C~\textsc{ii}] line. Based on global measurements of galaxies, \citet{Zanella2018-er} proposed the [C~\textsc{ii}] luminosity as a molecular gas mass tracer with a mass-to-light ratio $M_{\mathrm{gas}}/L_{\mathrm{[C~\textsc{ii}]}}=30M_{\odot}/L_{\odot}$.
However, our result suggests that the gas mass profiles estimated from [C~\textsc{ii}] do not agree with either CO- or dust-based profiles under the assumption of a spatially constant mass-to-light ratio for [C~\textsc{ii}]. Possible explanations for [C~\textsc{ii}] showing a flatter profile than CO and dust are (1) [C~\textsc{ii}] can be emitted from various gas phases including warm ionized gas and diffuse CO-dark gas, which are likely to extend to regions beyond those that can be traced by CO emission \citep{Pineda2013-wx}, (2) in the central region the C$^{+}$ abundance decreases because the high surface density shields gas from UV photons, leading most carbon to transition to CO \citep{Narayanan2017-mh}, and (3) the effect of infrared background radiation as proposed by \citet{Walter2022-uh}.

The CO-derived gas mass profile shows a steeper profile than the dust-derived one, especially in the central $\sim$1kpc. The excess in the centre may be due to the radially different excitation conditions, where the quasar radiation dominates in the central region and intense star formation dominates in the outer part of the galaxy \citep{Carilli2013-im}. In addition to the excitation conditions, some of the same mechanisms that flatten the [C~\textsc{ii}] profile may steepen the CO one: (1) the dust may trace not only molecular gas but also generally more extended atomic gas \citep{Orellana2017-dt}; (2) in the central region the CO abundance may be enhanced by increased shielding against dissociating UV photons \citep{Narayanan2017-mh}. However, in contrast to [C~\textsc{ii}], the infrared background radiation effect proposed by \citet{Walter2022-uh} is not significant for CO(7-6) because the dust continuum optical depth at the CO(7-6) frequency is $<0.1$. 

In addition to the normalized profiles shown in Fig.~\ref{fig:fig5}, we can also examine the absolute surface densities. If we derive the molecular gas mass profile using $\alpha_{\mathrm{CO}}=0.8$ M$_\odot$ pc$^{-2}$/ (K km s$^{-1}$) and $\mathrm{r}_{21}=L_{\mathrm{CO}(2 \rightarrow 1)}^{\prime}/ L_{\mathrm{CO}(1 \rightarrow 0)}^{\prime}=0.85$ as we did for our total molecular mass estimate, we automatically obtain the same estimate of the total molecular gas mass as \citet{Jones2016-cj}, $1.03\pm0.07\times 10^{11} M_\odot$. To match the CO-estimated and dust-estimated gas masses at 2kpc would require a gas-to-dust ratio 44 $\pm$ 24, with a relatively large error bar due to the large dispersion of the derived dust mass. This gas-to-dust mass ratio is consistent within the uncertainties with that derived from the whole-galaxy SED fit (see Fig.~\ref{fig:fig2}). Further exploration of gas emissivities per unit mass in various tracers is beyond the scope of this work, but this spatially resolved analysis \add{may} highlight that assumptions of a constant emissivity break down when appli\add{ed} to spatially-resolved data. (See Appendix C of \citealt{Herrera-Camus2021-fp} for further discussion of [C~\textsc{ii}] emissivity per unit mass in the optically thin limit and negligible background emission).

\subsection{The correlation of the temperature map and velocity dispersion map}\label{subsec:subsec35}
\begin{figure*}
\includegraphics[width=1.7\columnwidth]{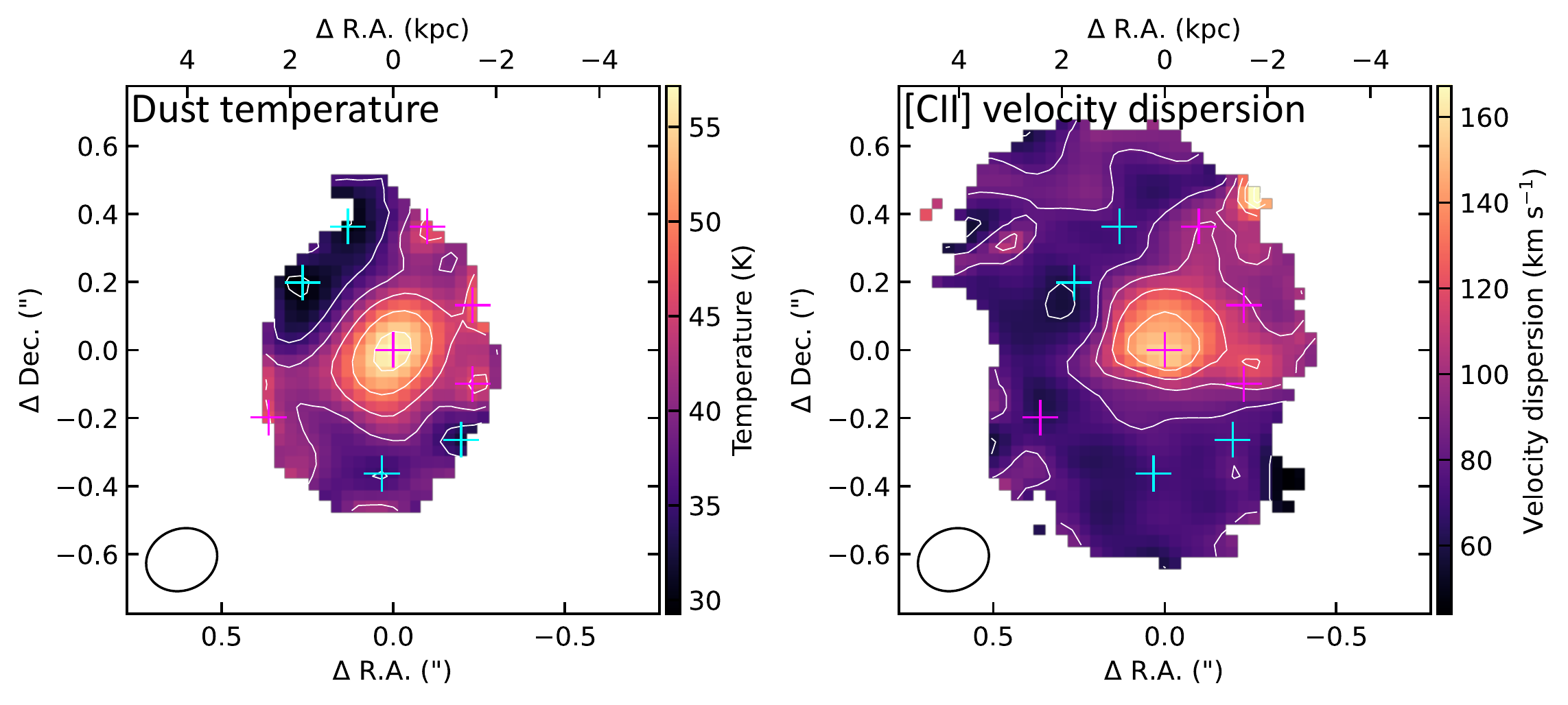}
\caption{The temperature map (left) and velocity dispersion map (right) as shown in Fig.~\ref{fig:fig3}A and Fig.~\ref{fig:fig1}F, respectively. The identified high- and low-temperature peaks (magenta and cyan crosses) are overlaid on the maps. The size of the synthesized beam (FWHM) is shown in the lower-left corner of each panel.
\label{fig:fig6}}
\end{figure*}

In Fig.~\ref{fig:fig6} we compare the dust temperature and velocity dispersion maps overlaid with low and high-temperature peaks. We see that high-temperature peaks coincide with regions of enhanced velocity dispersion extending from the centre to the north-west direction, while low-temperature peaks coincide with low velocity dispersion regions along the disk major axis. Such anisotropic distributions preferentially aligned with the disk minor axis may indicate that an outflow from the central region heats up the dust and enhances the velocity dispersion. In a nearby luminous infrared galaxy, NGC6240, \citet{Saito2017-mb} find a similar bipolar distribution for the high velocity dispersion CO emission, which coincides with the spatial distribution of H$\alpha$, near-IR, and X-ray emission. To examine the potential bulk gas motion due to the outflow which deviates from the mean line of sight motion in BRI 1335-0417, we fit the [C~\textsc{ii}] spectrum at each position of the high-temperature peaks. The spectra can be fitted with only a single Gaussian component and do not require any additional components given the current noise level of the data.

Another possible origin of the correlated enhancement of the temperature and gas velocity dispersion is energy injection by intense star formation. Using \add{adaptive optics} assisted Keck observations, \citet{Oliva-Altamirano2018-ha} suggested a spatial correlation between the peaks of the Pa$\alpha$ velocity dispersion and SFR for a relatively low redshift galaxy sample ($z=0.07$ to 0.2). If this scenario is correct, we expect high velocity dispersion, high-temperature regions to show high gas densities in order to host and sustain the intense star formation; these regions might include giant star-forming clumps or interacting satellites. Neither the derived dust mass surface density (see Fig.~\ref{fig:fig3}B) nor the observed [C~\textsc{ii}] and CO(7-6) maps (see Fig.~\ref{fig:fig1}) show such a distinct dense region, though they may show a broader triangle shape. However, \citet{Tadaki2018-au} identified 200pc-scale clumpy structures in a highly star-forming submillimeter galaxy at redshift $z=4$, both in the CO line and dust continuum. Whether similar small-scale clumpy structures occupy the temperature-enhanced regions in our target cannot be addressed with the current data due to sensitivity and resolution limits. To further constrain this scenario, higher-resolution data capable of tracing denser gas are required. 

\add{\subsection{Systematic error of the spatially resolved measurements}\label{subsec37}}

We confirm that the absolute flux uncertainty does not change the overall temperature distribution. Considering the worst possible case, when the true Band 7 flux and Band 9 flux are 10\% higher and lower than the observed values, respectively, the temperature would be underestimated by 2.7K in the median. In the \add{i}nverse scenario where the true Band 7 flux and Band 9 flux are 10\% lower and higher than the observed values, the temperature would be overestimated by 2.0K in the median. Therefore, the systematic error due to the absolute flux uncertainty is estimated to be +2.7K/$-2.0$K. The flux normalization errors only induce a pixel-to-pixel variation of 0.67 K and 0.49 K, respectively.

\add{We used the fiducial $\beta$ value $\beta=2.14\pm0.17$ derived from the integrated SED modelling, which is close to the recently measured values for the high redshift galaxies: median $\beta=1.9\pm0.4$ for ALESS well-constrained sub-samples \citep{Da_Cunha2021-ma} and $\beta=2.1\pm0.1$ for 47 starburst galaxies \citep[][reported as a private communication in \citealt{Da_Cunha2021-ma}]{Birkin2021-bh} both derived in the spatially integrated manner.} 

\add{However, the $\beta$ value can spatially vary over the galaxy, as shown in the nearby galaxies \citep{Galametz2012-yy, Smith2012-yd}. We investigate the effect on the derived dust temperature and mass profiles if the $\beta$ value changes across the galaxy from 1 to 2.5. Adopting a small $\beta$ of 1 increases the overall temperature by $\sim$20 K, while a large $\beta$ of 2.5 decreases the overall temperature by $\sim$2.5 K. This means a radially decreasing $\beta$ from 2.5 to 1.0 can erase the temperature gradient of 20 K seen in Fig.~\ref{fig:fig4}. However, the scenario is unlikely because it contradicts the generally observed anti-correlation of the temperature and $\beta$ in nearby galaxies \citep[warmer/colder dust has lower/higher $\beta$: e.g.][]{Galametz2012-yy, Smith2012-yd}. The expected radial increase in $\beta$ would increase rather than decrease the temperature gradient. The variation of $\beta$ also would not significantly change the mass profile. The induced changes in the dust mass profile are at most 40\%, which is smaller than the uncertainty of dust mass due to the dust opacity coefficient $\kappa_0$ \citep[at least a factor of 2, see, e.g.][]{Clark2019-sr}.}

As a final remark on the analysis in this section, note that the measured flux of an individual pixel represents the integrated value weighted by the beam, so the derived temperature and optical depth need to be considered as the luminosity weighted spatially averaged solutions over the region subtended by the beam ($\sim$\add{1.92} kpc$^2$). \add{It is expected that such spatial smoothing by the beam acts to flatten any intrinsic distribution.} 

As we will see in the next Section, there is a central unresolved component and an extended component in the dust continuum images, and the central unresolved component coincides with the high-temperature peak position. If the central unresolved component is due to the warm dust heated by the AGN, the derived high temperature is just due to the result of forcibly fitting two dust components with different temperatures with a single greybody function. This contamination is potentially significant out to the beam FWHM in radius. Similarly, we may overestimate the SFR in this area if we naively convert from our derived spatially resolved TIR luminosity to SFR, ignoring the possibility that the warm dust can contaminate our measurement. Therefore, it is important to decompose the image into the unresolved and the extended component and remove the effect of the unresolved component to correctly estimate the SFR distribution in the galaxy. We do exactly this in the next section.

\section{AGN-host galaxy decomposition}\label{sec:result2}
\subsection{Image decomposition of Band 7 and Band 9 continuum images}
In the previous section, we \add{demonstrated} that the dust temperature shows a steep increase up to $\sim$ 57K, which is much higher than the typical dust temperature assumed for quasar hosts \citep[47K;][]{Beelen2006-fc} and is presumably due to the heated dust by the central AGN. The observed dust continuum images (Fig.\ref{fig:fig1}) show a strongly peaked structure in the centre as well as a disky structure with spiral- and bar-like features in the outer part, suggesting the presence of a distinct central unresolved component and a resolved extended component. The former is likely to be associated with the warm dust heated by the central AGN and the latter with the cold dust heated by star formation in the host galaxy.

For these reasons, we decompose the continuum image into an unresolved AGN component and an extended host galaxy component using the observed [C~\textsc{ii}] moment0 map as a template tracing the dust distribution heated by the star formation in the galaxy. This approach has several advantages: (1) there is only one parameter (total flux), (2) the [C~\textsc{ii}] emission captures disk substructures like arms and a bar that are also seen in the dust continuum map, and that are not easily representable by an analytic model, and (3) there is independent evidence that [C~\textsc{ii}] is a good overall neutral gas tracer \citep[e.g.,][]{Herrera-Camus2018-vf, Herrera-Camus2018-cx}; we expect this gas to form stars at a rate roughly proportional to the density \citep{Krumholz2011-km}, and there is good evidence that [C~\textsc{ii}] directly correlates with star formation \citep{De_Looze2011-jm}. [C~\textsc{ii}] in this galaxy also has rotating disk kinematics and a nearly exponential profile typical of disks (Sersic index $n\sim0.9$) \citep{tsukui2021-vl}. Given these considerations, our decomposition model includes the following components: (1) a component proportional to the point spread function \add{(PSF)} of the observation $I_{\mathrm{PSF}}$ to reproduce the unresolved emission, (2) a component with the same spatial distribution as the observed [C~\textsc{ii}] intensity map representing a disky, star-forming component that exhibits spiral and bar-like structure, $I_{\mathrm{disk}}$, and (3) a circular Gaussian component, convolved by the \add{PSF} of the observation, $I_{\mathrm{Gauss}}$, which we include in case the central compact region includes emission that is extended rather than truly point-like (e.g., a nuclear starburst), and thus is partially resolved by the ALMA beam. The model with these three components is described as

\begin{equation}
\begin{aligned}
I_{\mathrm{model}}= & I_{\mathrm{PSF}}(cx,cy, f_{\nu,\mathrm{point}}) \\
                    & +I_{\mathrm{Gauss}}(cx,cy, \sigma_{\mathrm{Gauss}}, f_{\nu,\mathrm{Gauss}}) \\ 
                    & +I_{\mathrm{disk}}(f_{\nu,\mathrm{disk}}),
\end{aligned}
\end{equation}
where the model has only 6 parameters: the central coordinates of the PSF and Gaussian components ($cx, cy$), the intrinsic size of the Gaussian $\sigma_{\mathrm{Gauss}}$ (standard deviation), and the total flux of each component $f_{\nu,\mathrm{point}}$, $f_{\nu,\mathrm{Gauss}}$, and $f_{\nu,\mathrm{disk}}$. We create the spatial template for the disk component $I_{\mathrm{disk}}$ from the [C~\textsc{ii}] moment map as follows: first, we convolve the [C~\textsc{ii}] cube to the same resolution as the observed continuum image we are fitting, and then we make a [C~\textsc{ii}] moment map by the masked moment method (\citealt{Dame2011-nv}; convolution with a kernel of 2 times the beam size and 2 times the spectral channel width, then masking with a threshold set to the rms of the original cube) to reduce the bias introduced by the presence of noise in the [C~\textsc{ii}] moment map. This bias pushes the best-fit model toward lower $f_{\nu,\mathrm{disk}}$ values because larger $f_{\nu,\mathrm{disk}}$ values amplify the noise present in the [C~\textsc{ii}] moment map, giving an additional penalty in $\chi^2$; convolution and masking suppress this noise in the spatial template and therefore suppress the bias.

\begin{figure*}
\centering
\includegraphics[width=\textwidth]{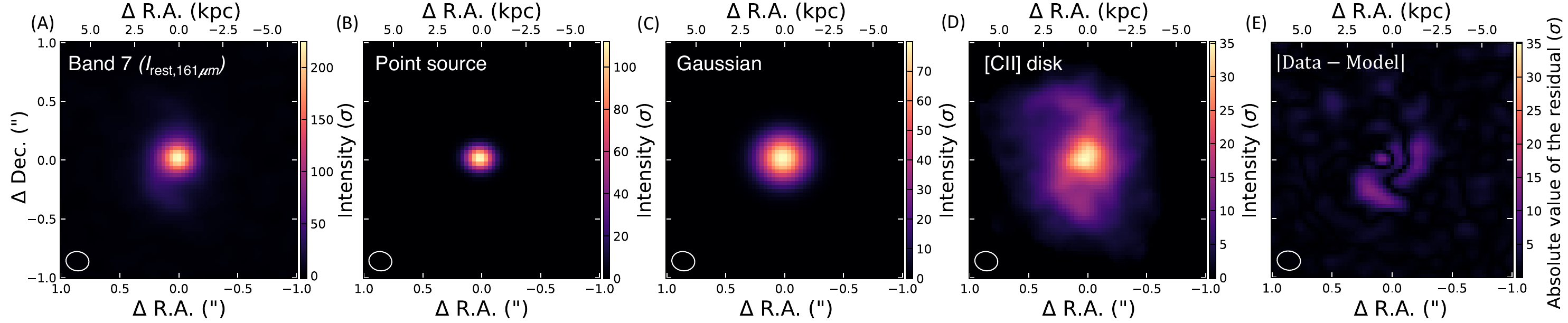}
\caption{Decomposition of the rest-frame $161~\mu\mathrm{m}$ continuum image: (A) the original image on a linear scale; (B) the best-fit point source component; (C) the best-fit circular Gaussian component; (D) the best-fit [C\textsc{ii}] component; (E) the \add{absolute} residual \add{|}data $-$ model\add{|}. The intensity scale for all maps is in units of the rms noise $\sigma$ of the continuum image (A). We show the size of the synthesized beam (FWHM) in the lower-left corner of each panel.\label{fig:fig7}}
\end{figure*}

\begin{figure*}
\centering
\includegraphics[width=\textwidth]{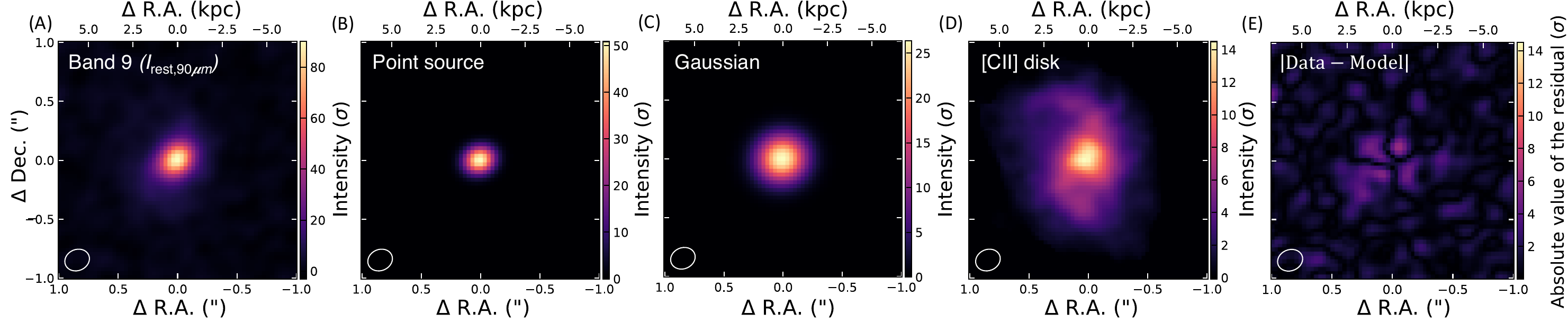}
\caption{Decomposition of the rest-frame $90~\mu\mathrm{m}$ continuum image. Same as in Fig.~\ref{fig:fig7} except showing Band 9 results. \label{fig:fig8}}
\end{figure*}

We fit $I_{\mathrm{model}}$ to the observed Band 7 and Band 9 continuum images with simple $\chi^2$ minimization. For both the Band 7 image $I_{\lambda_{\mathrm{rest}}=161~\mu\mathrm{m}}$ and the Band 9 image $I_{\lambda_{\mathrm{rest}}=90~\mu\mathrm{m}}$, evaluation of the Bayesian information criterion (BIC) indicates that this three-component fit is preferred over any alternatives omitting one of the components. Full details are provided in Appendix [see Fig.~\ref{fig:figa2} and Fig.~\ref{fig:figa3} for BIC and chi-square values, see Fig.~\ref{fig:figa4} and Fig.~\ref{fig:figa5} for fitting residuals]. Figures~\ref{fig:fig7} and \ref{fig:fig8} show the original image (A: \add{see Fig.~\ref{fig:fig1} for the same images in log scale}), the best-fit PSF model (B), the Gaussian model (C) and the [C~\textsc{ii}] model (D), the best-fit residual (E) for Band 7 and Band 9 continuum images, respectively.

In Table~\ref{tab:tab2}, we show the best-fit parameters we derive. \add{The uncertainties on these parameters are computed} by Monte Carlo resampling with noise which has realistic spatial correlations. Our procedure to make this estimate is to: (1) measure the autocorrelation function (ACF) of the noise map in the primary beam uncorrected Band 7 and Band 9 continuum images, (2) randomly generate noise maps with the same correlation properties characterized by the noise ACF using the ESSENCE package \citep{Tsukui2022-ne, 2023arXiv230103579T}, (3) repeat the fitting procedure 300 times with different random realizations of the noise. We then take the 68\% confidence interval on the resulting distribution of parameters as our confidence interval in Table~\ref{tab:tab2}. 

Our fits pass several consistency checks. First, the measured total flux $f_{\nu,\mathrm{total}}$ from the observed images and the model total flux $f_{\nu,\mathrm{model}}=f_{\nu,\mathrm{point}}+f_{\nu,\mathrm{Gauss}}+f_{\nu,\mathrm{disk}}$ agree within the statistical uncertainty for both images. Second, the centre position of the point source and Gaussian components coincides with the location of the optical quasar position and the highest peak in the continuum images and temperature map within the uncertainties. Third, the sizes of the additional Gaussian components that we derive independently by fitting the two continuum images are consistent within the uncertainty. We speculate that this component is likely needed to compensate for the central decrease in the [C\textsc{ii}]/FIR mentioned in Section \ref{sec:result}. We do see some correlated residuals in the residual map of $I_{\lambda_{\mathrm{rest}}=161~\mu\mathrm{m}}$ (Fig.~\ref{fig:fig7}), which shows a positive residual at the location of the southern arm but not the northern arm. This indicates that dust continuum emission is enhanced relative to [C~\textsc{ii}] emission in the southern arm but not in the northern arm. The residual of $I_{\lambda_{\mathrm{rest}}=90~\mu\mathrm{m}}$ also shows a positive residual in a similar position, but without a clear spiral structure (Fig.~\ref{fig:fig8}). This result may be interpreted as the two arms having different ISM properties (opacity, temperature, metallicity, etc.). However, despite this feature, our simple model with only 6 parameters describes well the overall morphology of the continuum emission without any other significant structures in the residual. 

\begin{figure*}
\centering
\includegraphics[height=0.35\textheight]{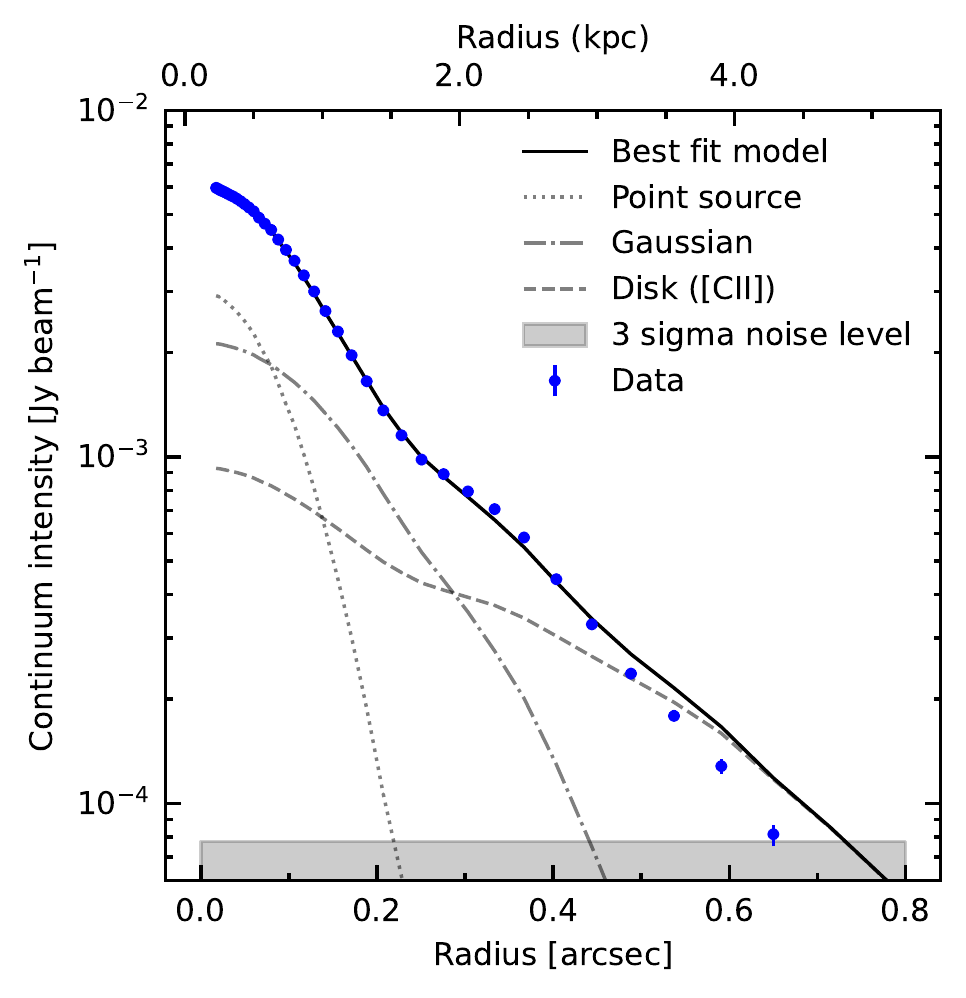}
\includegraphics[height=0.34\textheight]{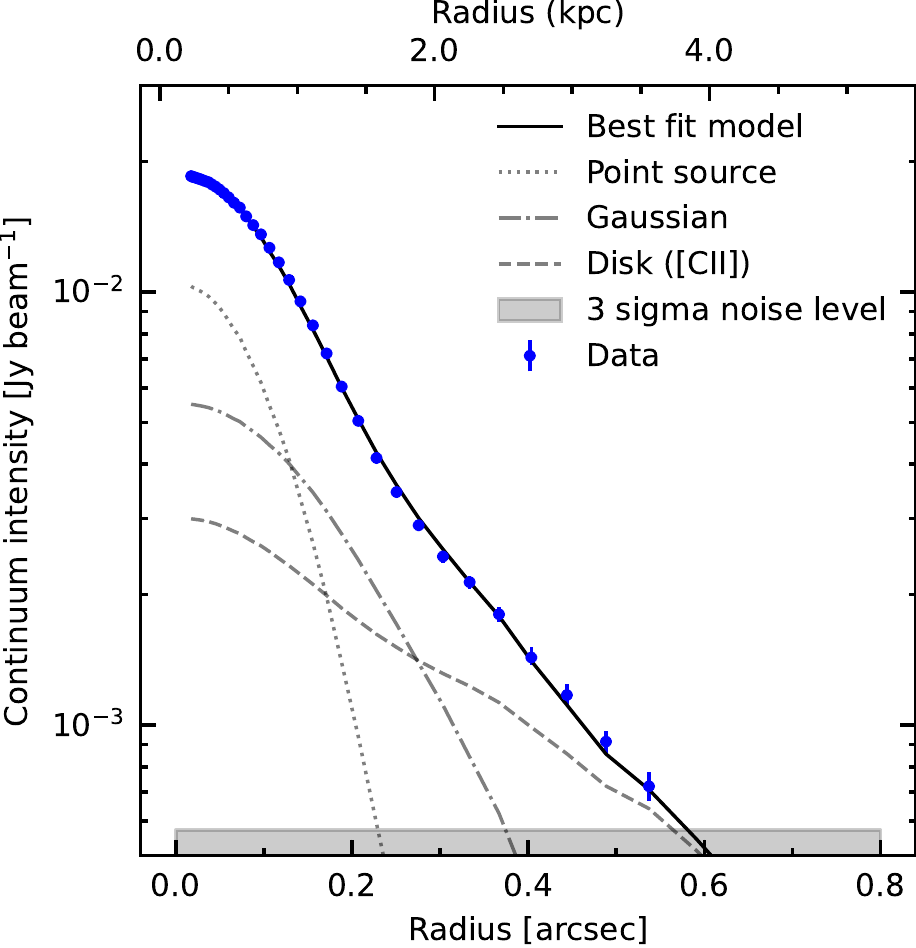}
\caption{The surface brightness profiles for rest-frame $161~\mu\mathrm{m}$ (left) and $90~\mu\mathrm{m}$ (right) continuum images. 
Blue points with an error bar show the azimuthally-averaged surface brightness profile of the continuum image, while lines show the best-fit decomposed model, including the point source (dotted grey line), Gaussian (dot-dashed grey line), and [C\textsc{ii}] (dashed grey line) components, and their sum (black solid line). These profiles are derived using Photutils code \citep{Larry_Bradley2019-lk} as follows. We first fit elliptical isophotes to the dust continuum images, where the four parameters describing each ellipse (2-dimensional centre position, position angle, and ellipticity) are left free. We then measure azimuthally-averaged surface brightness along each of the best-fitting ellipses. The error bars on the data points include uncertainties in the ellipse fitting due to spatially correlated noise in the images. \label{fig:fig9}}
\end{figure*}

Figure~\ref{fig:fig9} shows the surface brightness profile of the Band 7 and Band 9 continuum images and their best-fit decomposed components. The contribution of the point sources to the integrated flux over the image is small, 15\% and 21\% in rest-frame $161~\mu\mathrm{m}$ and rest-frame $90~\mu\mathrm{m}$, respectively (Table~\ref{tab:tab2}). However, the point source dominates the flux in the central pixel (flux in the central resolution element, or beam area), contributing 49.8$_{-1.3}^{+1.5}$\% and 56.7$^{-6.1}_{+4.5}$\% in $I_{\lambda_{\mathrm{rest}}=161~\mu\mathrm{m}}$ and $I_{\lambda_{\mathrm{rest}}=90~\mu\mathrm{m}}$, respectively. The best-fit model profile is slightly higher than the dust continuum profile at the outskirts of the galaxy ($\sim$ 0.6 arcsec, Fig.~\ref{fig:fig9} (left)). This indicates the [C~\textsc{ii}] model component is more extended than the FIR emission, even after the central compact point source and the compact Gaussian component are removed. This is consistent with the presence of extended [C~\textsc{ii}] halo structures extending farther than the underlying FIR emission (out to $\sim$10kpc) in other star-forming galaxies at a similar redshift (e.g., \citealt{Fujimoto2019-vw, Fujimoto2020-ff}; however for a contrary view see \citealt{Novak2020-yi}, who find no evidence for halo structures in quasar host galaxies at z$>$6 using stacking analysis). The relatively short (1h) Band 7 observation relying on the brightness of the source is not sensitive enough to probe the faint structure out to 10 kpc in BRI 1335-0417. Note that in what follows, we use only the point source fluxes $f_{\nu,\mathrm{point}}$ and the measured total fluxes $f_{\nu,\mathrm{total}}$ from the images, not the fluxes of the individual Gaussian $f_{\nu,\mathrm{Gauss}}$ or [C~\textsc{ii}] disk components $f_{\nu,\mathrm{disk}}$, or the total model fluxes $f_{\nu,\mathrm{model}}$. Therefore, the slight model-data offset in the surface brightness profile at the outer part of the galaxy does not affect our subsequent results.

\begin{table*}
    \centering
    {\tabulinesep=1.1mm
    \begin{tabu}{c c c}
         \hline\hline
         Parameters & Band 7 ($\lambda_{\mathrm{rest}}=161\micron$) & Band 9 ($\lambda_{\mathrm{rest}}=90\micron$)\\
         \hline\hline
         x offset of the centre (R.A), $cx$ (mas) & 19.6$\pm0.5$ & 9.6$\pm1.6$\\
         y offset of the centre (Dec.), $cy$ (mas) & 13.6$\pm0.6$ & 1.3$\pm1.4$\\
         Point source flux, $f_{\nu,\mathrm{point}}$ (mJy)& 3.09$^{+0.09}_{-0.08}$ & 10.87$^{+0.85}_{-1.18}$\\
         Gaussian flux $f_{\nu,\mathrm{Gauss}}$ (mJy)& 7.4$^{+0.18}_{-0.14}$ & 16.83$^{+1.08}_{-1.06}$\\
         \text{[C\textsc{ii}]} template flux $f_{\nu,\mathrm{disk}}$ (mJy)& 10.26$^{+0.25}_{-0.36}$ & 27.62$^{+2.32}_{-2.19}$\\
         Gaussian size $\sigma_{\mathrm{Gauss}}$ (pc)& 787$^{+18}_{-17}$ & 777$^{+66}_{-75}$\\
         Model total flux $f_{\nu,\mathrm{model}}$ (mJy)& 20.76$\pm$0.36 & 55.32$\pm$2.69\\
         \hline
         Derived parameters\\
         \hline
         Measured total flux $f_{\nu,\mathrm{total}}$ (mJy)& 20.54$\pm$0.27& 52.92$\pm$2.13\\
         \hline\hline
    \end{tabu}}
    \caption{The best-fit result of the image decomposition. The x and y offset of the centre of the point source and Gaussian is denoted relative to the image centre coordinate (RA, Dec)=($204.514231$deg, $-4.543055$deg), which corresponds to the highest peak pixel of the temperature. The offset of the quasar position from the centre is RA$=14.5$mas and Dec$=4.5$mas, consistent with that of the point source component within the ALMA's positional uncertainty. Model total flux $f_{\nu,\mathrm{model}}$ is the sum of each component flux $f_{\nu,\mathrm{point}}+f_{\nu,\mathrm{Gauss}}+f_{\nu,\mathrm{disk}}$. Measured total flux $f_{\nu,\mathrm{total}}$ are measured from the images, same as listed in Table~\ref{tab:taba1}.}
    \label{tab:tab2}
\end{table*}

\subsection{SED fitting using the image decomposition result}
In the previous section, we found that an unresolved component is required to reproduce the dust continuum images, and we decompose the images into a sum of this component and the rest of the galaxy. In this subsection, we model the SEDs of the unresolved component (AGN-heated dust), with flux $f_{\nu, \mathrm{point}}$, and the rest of the galaxy, with flux $f_{\nu, \mathrm{total}}-f_{\nu, \mathrm{point}}$ (cold\add{er} dust heated by the star formation in the disk). We use the fluxes of the \add{point-like} components $f_{\lambda_{\mathrm{rest}}=161\mu\mathrm{m}, \mathrm{point}}$ and $f_{\lambda_{\mathrm{rest}}=90\mu\mathrm{m}, \mathrm{point}}$ to constrain the AGN-heated warm dust contribution to the FIR part of SED, along with total flux measurements at various wavelength bands listed in Table~\ref{tab:taba1}. Note that we do not use the fluxes of the rest of the galaxy $f_{\nu, \mathrm{total}}-f_{\nu, \mathrm{point}}$ as independent data points, since they are already constrained implicitly by the total flux measurements $f_{\nu, \mathrm{total}}$ and point source flux measurements $f_{\nu, \mathrm{point}}$. We consider several possible approaches to fitting the resolved and unresolved components described below.

\subsubsection{\add{Two greybody components SED modelling}}
We first model the FIR part of the SED with two greybody functions in the optically thin limit (Eq.~\ref{eq:eq1}, $\tau \ll 1$), one for the AGN-heated dust and the other for the cold dust heated by star formation. Although detailed AGN dust torus models exist to describe the warm dust component heated by AGN \citep{Honig2017-sv, Stalevski2016-em}, the spatial extent and geometry of tori are highly uncertain for high redshift quasars. This motivates us to consider a simpler model where we fit the AGN-dust component with a greybody spectrum in the optically thin limit but with a free dust emissivity index $\beta$. 
This model has 6 free parameters ($T_\mathrm{dust}$, $M_\mathrm{dust}$, $\beta_{\mathrm{dust}}$ for each component). \add{The sum of the two greybody functions (AGN-heated dust and colder dust heated by star formation) is fitted to the spatially integrated fluxes (from rest-frame 36 to 472~$\micron$ bands in Table~\ref{tab:taba1}), and the AGN heated dust greybody is fitted to point fluxes, $f_{\lambda_{\mathrm{rest}}=161\mu\mathrm{m}, \mathrm{point}}$ and $f_{\lambda_{\mathrm{rest}}=90\mu\mathrm{m}, \mathrm{point}}$ (Table~\ref{tab:tab2}). The fitting is done simultaneously to constrain both the AGN-heated dust component and the colder dust component.} 

\begin{figure*}
\centering
\includegraphics[width=0.6\textwidth]{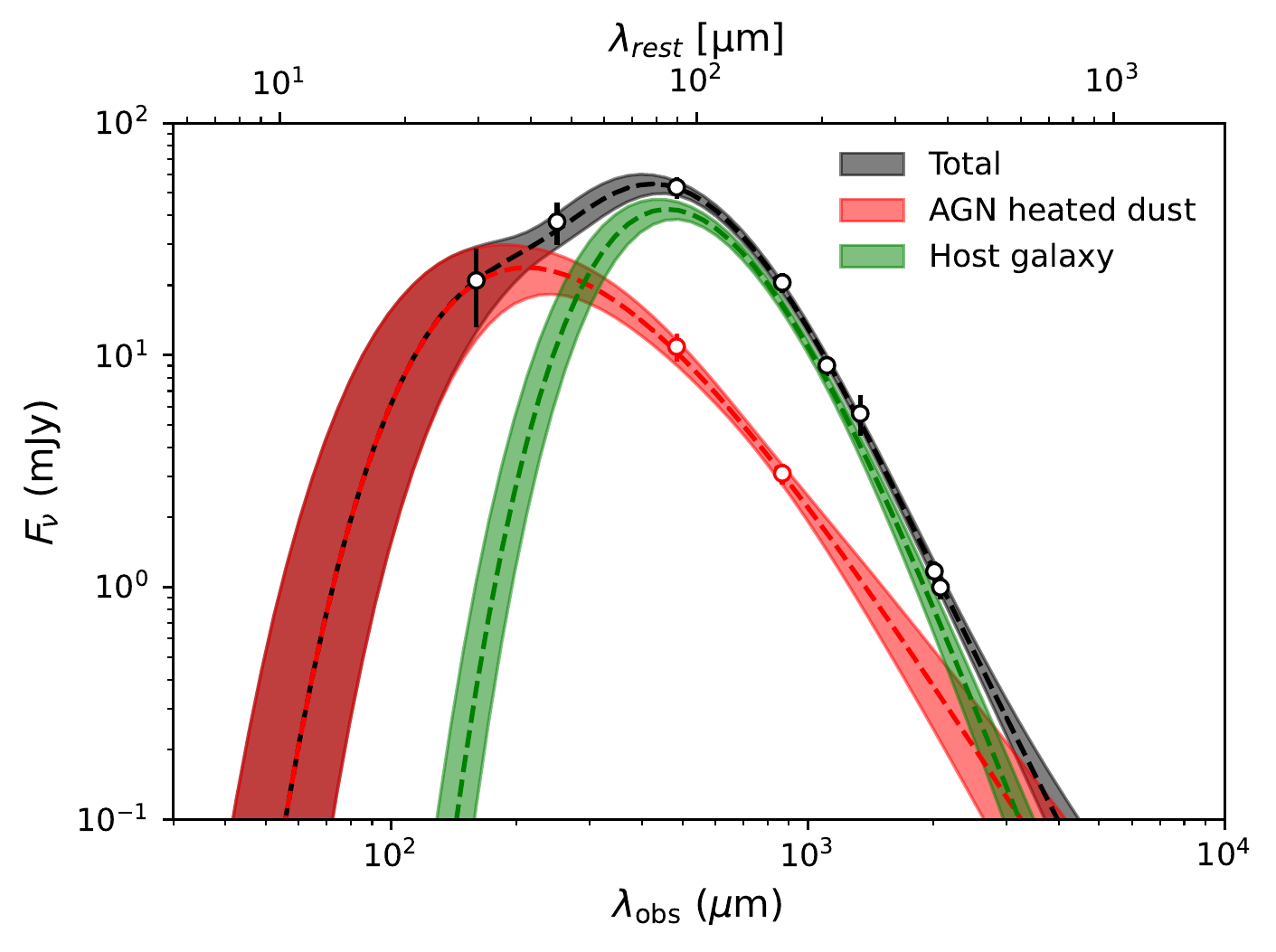}
\caption{The far infrared spectral energy distribution of BRI1335-0417 (black points, see Table~\ref{tab:taba1}) with the best-fit model (grey) composed of two greybody spectra: a warm dust component heated by the AGN (red) and a cold dust component associated with the host galaxy (green). Both components are assumed to be optically thin. In addition to the total photometric points, we used point source fluxes, $f_{\lambda_{\mathrm{rest}}=161\mu\mathrm{m}, \mathrm{point}}$ and $f_{\lambda_{\mathrm{rest}}=90\mu\mathrm{m}, \mathrm{point}}$ (red points) measured by the image decomposition to constrain the flux contributed by the AGN-heated warm dust component (Table~\ref{tab:tab2}). \label{fig:fig10}}
\end{figure*}

We compute the distribution of their posteriors for a uniform prior distribution using the \textsc{emcee} package \citep{Foreman-Mackey2013-yn}. Figure~\ref{fig:fig10} shows the SED decomposed into the AGN-heated dust component and the host galaxy. Figure~\ref{fig:figa6} shows the posterior distribution of the model parameters, indicating that all model parameters are well-constrained. From this fit, we find that the SFR of the host galaxy is 1.5$^{+0.3}_{-0.2}\times 10^{3}M_\odot$ yr$^{-1}$ based on the FIR luminosity of the cold dust component, which is more than 3 times smaller than the previous estimate \add{$5040\pm{1300} M_\odot$ yr$^{-1}$} \citep{Wagg2014-bh}. This is due to the high AGN contribution, 63$^{+15}_{-16}$\%, to the FIR luminosity, which is neglected in the previous study. The AGN-heated dust component has a temperature $T_{\mathrm{dust}}=95.60^{+37.33}_{-25.81}$ K, which is consistent with the high temperature reported for the centre of high redshift quasars found in earlier high-resolution observations ($\sim$200 pc; \citealt{Walter2022-uh, Shao2022-ft}). However, from this fit, we also find that the temperature of the cold dust component is $T_{\mathrm{dust}}=29.3^{+4.5}_{-3.6}$ K, which is lower than the temperature $29-47$K with a median value of $\sim38$ K at the outer part of the galaxy (Fig.~\ref{fig:fig4}). This is expected due to the optically thin assumption we adopted in the SED modelling, which yields systematically lower temperatures than fits that allow the dust to have finite optical depth \citep{Cortzen2020-dq}. 

Improving this aspect of our fit motivates our second fitting method, wherein we modify our model by introducing an additional free parameter for the dust optical depth $\tau_\nu$ at $161\micron$, which we denote $\tau_{161\mu \mathrm{m}}$, for the cold dust component. We fit the model again with the same procedure for the above optically thin case, but with a prior constraint that the cold dust temperature cannot exceed the warm dust temperature. Figure~\ref{fig:figa7} shows the posterior distribution of the model parameters, which are summarized in Table~\ref{tab:tab3}.
\footnote{Note that, since the posterior distribution for $\tau_{161\mu \mathrm{m}}$ remains significantly above zero all the way down to $\tau_{161\mu \mathrm{m}}=0$, for this parameter report only a best-fit value and an upper limit $\tau_{161\mu \mathrm{m}}<0.0146$ (84th percentile of the posterior distribution), not a lower limit.} Including a finite optical depth increases our best-fitting cold dust temperature to $52.6_{-11.0}^{+10.3}$ K. This confirms that relaxing the optically thin assumption provides a dust temperature consistent with the spatially resolved result, while at the same time not significantly changing the overall functional shape of two greybody functions or the parameters derived from them such as the SFR and the AGN luminosity. The updated SED fit is shown in Fig.~\ref{fig:figa8}. The derived range of optical depth $\tau_{161\mu \mathrm{m}}$ from 0 to 0.0146 is much smaller than the spatially resolved result with the median value of 0.1556. The optical depth derived by the spatially integrated SED assuming single temperature $T_\mathrm{dust}$ may have much less physical meaning than the optical depth derived in the spatially resolved image (Fig.~\ref{fig:fig3}B) because the galaxy shows a significant variation of non-linear parameters such as temperature and optical depth over the galaxy. This may illustrate that the optical depth $\tau_{\nu}$ measurement becomes inaccurate if the dust temperature and opacity structure are smaller than the integration aperture. 

From this fit we also obtain dust emissivity indices $\beta_{\mathrm{dust}}=0.91^{+0.56}_{-0.49}$ and $\beta_{\mathrm{dust}}=2.56^{+0.37}_{-0.30}$ for the warm and cold components, respectively. Our finding of a shallow index $\beta\sim 1$ for the warm component and a steeper index $\beta\sim 2.5$ for the cool component is consistent with the large range of $1\lesssim \beta \lesssim 2.5$ suggested by the observations and models of infrared to sub-millimeter dust emission and laboratory experiments (e.g., laboratory experiments: \citealt{Mennella1998-iw, Agladze1996-ny}, models; \citealt{Pollack1994-tq}, observations: \citealt{Galametz2012-yy, Dupac2003-pd}). In agreement with our results, these studies suggest that $\beta_{\mathrm{dust}}$ anti-correlates with the dust temperature $T_{\mathrm{dust}}$; $\beta \sim1$ for small grains, primarily radiating in MIR bands, and $\beta\sim2$ for large grains, which can \add{achieve} lower equilibrium temperature and thus radiate their energy in FIR bands \citep{Da_Cunha2008-jt}.

\subsubsection{\add{Stardust panchromatic SED modelling}}
\begin{figure*}
\centering
\includegraphics[width=0.7\textwidth]{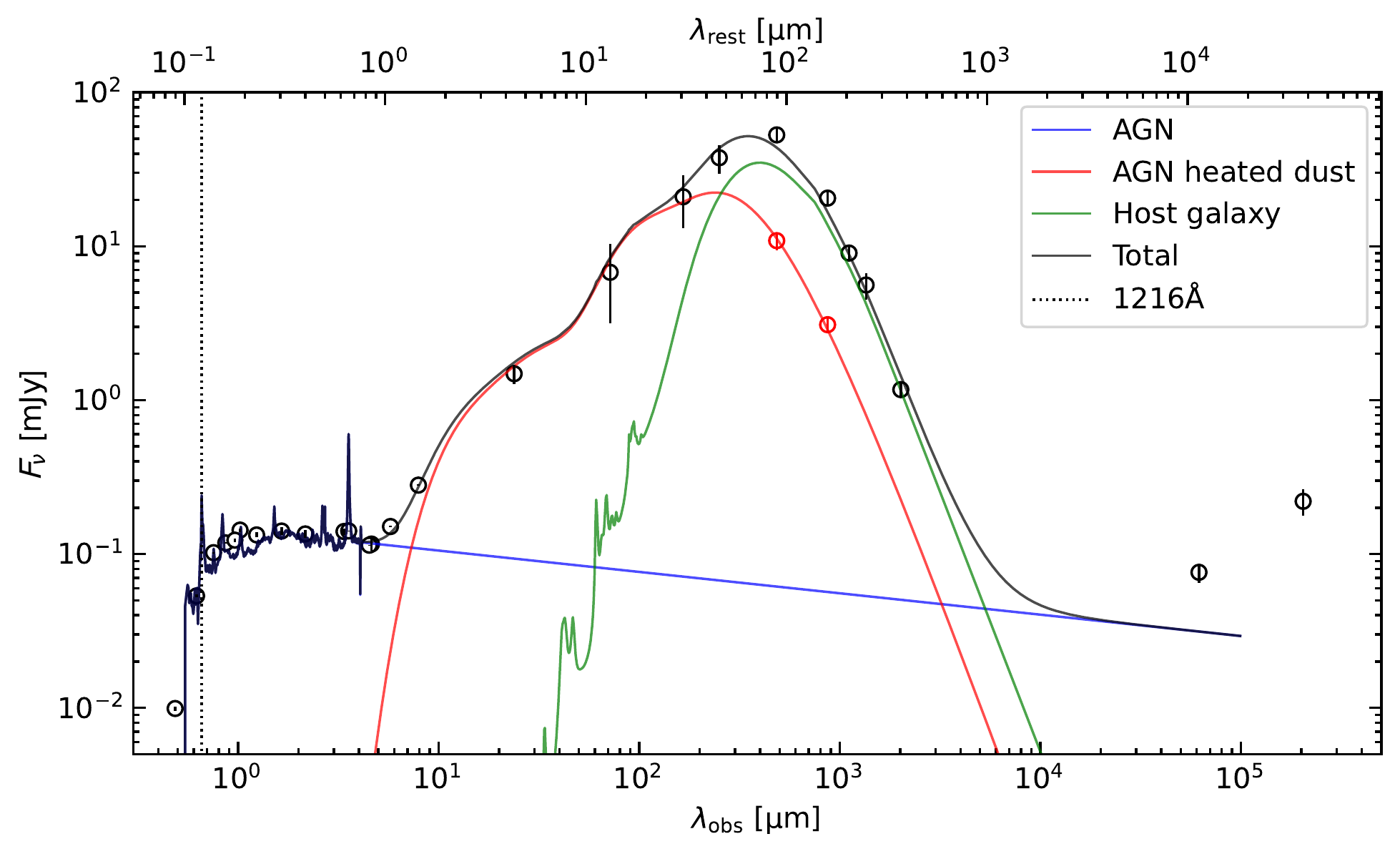}
\caption{The UV to Radio spectral energy distribution of BRI1335-0417 (black points, see Table~\ref{tab:taba1}) with the best-fit \textsc{stardust} model \citep{Kokorev2021-yp} composed of a quasar template (blue solid line: UV to optical), AGN-heated hot and warm dust template (red solid line: near-IR to mid-IR), and cold dust associated with the host galaxy (green solid line: FIR). In addition to the total photometric points, we used point source fluxes, $f_{\lambda_{\mathrm{rest}}=161\mu\mathrm{m}, \mathrm{point}}$ and $f_{\lambda_{\mathrm{rest}}=90\mu\mathrm{m}, \mathrm{point}}$ (red points) measured by the image decomposition to constrain the flux contributed by the AGN-heated warm dust component (Table~\ref{tab:tab2}). \add{The dotted vertical line denotes the Ly$\alpha$ line wavelength.} \label{fig:fig11}}
\end{figure*}

Our third fitting approach is to model the panchromatic SED from the rest UV to FIR bands using the more realistic galaxy SED fitting code \textsc{stardust} \citep{Kokorev2021-yp}, which assumes three components; a QSO template dominating in the rest-frame UV to optical bands \citep{2016ApJ...817...55S}, an AGN-heated dust component \citep{Mullaney2011-sn} dominating in the rest-frame MIR bands, and a cold dust component of the host galaxy dominating in the FIR bands \citep{Draine2007-ou}. \add{The original \textsc{stardust} code fits the combination of the above three templates to the spatially integrated photometric data points given as input. In addition to the spatially integrated data, we modified the code to accept the point source fluxes as independent constraints to be fitted with the AGN-heated dust component. Using the modified code, we simultaneously fit (1) the sum of three templates to the spatially integrated data and (2) the AGN-heated dust component to point source fluxes, $f_{\lambda_{\mathrm{rest}}=161\mu\mathrm{m}, \mathrm{point}}$ and $f_{\lambda_{\mathrm{rest}}=90\mu\mathrm{m}, \mathrm{point}}$}
\footnote{The modified \textsc{stardust} code will be available at the GitHub repository \url{https://github.com/takafumi291}, upon publication}. Figure~\ref{fig:fig11} shows the best-fit SED decomposed into the QSO component (AGN-heated dust) and the host galaxy component (cold dust associated with the star-forming host galaxy). The SFR and AGN contribution to $L_{\mathrm{TIR}}$ we derive from this fit are consistent with those we obtain from the two-component greybody fit (Table~\ref{tab:tab3}). Our \textsc{stardust} calculation also finds that a QSO template without dust extinction ($A_{\mathrm{V}}=0$) fits the data better than a template including extinction, which may indicate that the QSO has already cleared dust at least from the line of sight between us and the central accretion disk. We also consider a fit from \textsc{stardust} using a stellar spectrum template instead of a quasar template for the UV-to-optical part of SED, but this results in a $\chi^2$ value that is larger by 536, with \add{16} degrees of freedom. This difference in $\chi^2$ is highly significant and strongly favours the quasar template. \add{This result is also supported by the \textit{HST} optical image (Fig. \ref{fig:fig0}) consistent with the point spread function.}
Moreover, the stellar template fit produces an estimated stellar mass of $\sim10^{12}M_\odot$, which easily exceeds the dynamical mass. This confirms that the UV to optical part of the SED is dominated by AGN emission, supporting our interpretation that the compact dust component identified at the position of the highest temperature peak is predominantly heated by AGN.

For completeness, we also perform \textsc{stardust} SED fitting \textit{without} using the image decomposition results, $f_{\lambda_{\mathrm{rest}}=161\mu\mathrm{m}, \mathrm{point}}$ and $f_{\lambda_{\mathrm{rest}}=90\mu\mathrm{m}, \mathrm{point}}$, to constrain the AGN-heated dust contribution to the FIR SED. In this case, \textsc{stardust} provides a best-fit model with a \add{AGN subtracted} SFR of 5260$\pm31M_{\odot}$, much higher than that derived when including the spatially-resolved constraints; moreover, the $\chi^2$ value of this fit ($\chi^2=282$, DOF=18) is no worse than that derived including the spatially resolved data. We show the fit in Fig.\ref{fig:figa9}, and visual inspection confirms that the fit to the global SED is no worse than the one shown in Fig.~\ref{fig:fig11} derived using the spatially decomposed constraints. This demonstrates that spatially resolved information at rest-frame $\sim90$ to $161\micron$ band is required to constrain and remove the AGN-heated dust component in order to estimate SFR accurately; it is not possible to make this correction with unresolved photometric data alone.

\begin{table*}
    \centering
    \begin{tabular}{c c c c c c}
         \hline\hline
          Parameters & unit & Two greybodies & Two greybodies& Stardust \\
          cold dust opacity assumption & & thin & thick &  thin\\
          \hline\hline
          $\log_{10}M_{\mathrm{dust}}$(warm dust) & ($M_{\odot}$) & $7.98_{-0.13}^{+0.13}$ & $8.02_{-0.13}^{+0.13}$ &\\
          $T_{\mathrm{dust}}$(warm dust) & (K) & $96.0_{-25.9}^{+38.9}$ & $87.1_{-18.3}^{+34.1}$ &\\
          $\beta_{\mathrm{dust}}$ (warm dust) & & $0.85_{-0.49}^{+0.56}$ & $0.91_{-0.49}^{+0.49}$ &\\
          $\log_{10}M_{\mathrm{dust}}$(cold dust) & ($M_{\odot}$) & $9.48_{-0.15}^{+0.14}$ & $9.00_{-0.13}^{+0.16}$ & $9.39\pm0.03$\\
          $T_{\mathrm{dust}}$(cold dust) & (K) & $29.3_{-3.6}^{+4.5}$ & $52.6_{-11.0}^{+10.3}$ & $37.8\pm0.6^\mathrm{a}$\\
          $\beta_{\mathrm{dust}}$ (cold dust) & & $2.75_{-0.40}^{+0.46}$& $2.56_{-0.30}^{+0.37}$ &\\
          $\tau_{\lambda_{\mathrm{rest}}=161\mu \mathrm{m}}$ & & & 0.0072 ($<0.0146$)& \\
          \hline
          Derived parameters & & & & \\
          \hline
          Total $L_{\mathrm{IR}}$ & ($10^{13} L_\odot$) & $3.9_{-0.8}^{+1.3}$ &  $3.8_{-0.6}^{+0.9}$ & $5.4\pm{0.1}$ &\\
          AGN $L_{\mathrm{IR}}$ & ($10^{13} L_\odot$) & $2.4_{-0.9}^{+1.3}$ & $2.0_{-0.7}^{+1.1}$ &$3.6\pm{0.1}$ & \\
          Host galaxy $L_{\mathrm{IR}}$ & $10^{13} L_\odot$ & $1.5_{-0.2}^{+0.3}$ & $1.7_{-0.4}^{+0.5}$ & $1.77\pm{0.02}$ &\\
          SFR & ($10^3 M_\odot$ yr$^{-1}$) & $1.5_{-0.2}^{+0.3}$ & $1.7_{-0.4}^{+0.5}$ & $1.77\pm{0.02}$ &\\
          AGN fraction in $L_{\mathrm{IR}}$ &  & $0.62_{-0.13}^{+0.11}$ & $0.53_{-0.15}^{+0.14}$ & $0.67\pm{0.01}$ &\\
          \hline\hline
          $\chi^{2}$ (DOF) & & 0.98 (4) & 0.66 (3) & 284.19 (2\add{0})\\
         \hline\hline
    \end{tabular}
    \caption{The derived parameter with SED modelling. $^{\mathrm{a}}$The stardust code also returns the average radiation field intensity $\left<U\right>=66.0\pm6.4$ ($= L_{\mathrm{dust}}/M_{\mathrm{dust}}/125$), which is a proxy of the luminosity-weighted dust temperature $T_\mathrm{dust}$. The temperature is converted by using the $\left<U\right>=(T_{\mathrm{dust}}/18.9K)^{6.04}$ following \citet{Magdis2017A&A}.}
    \label{tab:tab3}
\end{table*}

\subsubsection{\add{Conclusion of the SED fitting results}}
The most important conclusion from the discussion above is that all three methods of SED modelling that include the spatially-resolved constraints provide consistent estimates of the AGN contribution to $L_{\mathrm{TIR}}$, and therefore for the host galaxy SFR \add{(Table~\ref{tab:tab3})}. 
As final values for the remainder of this paper, we adopt the best-fit model parameters derived from the two greybodies fit with a free optical thickness $\tau_{\lambda_{\mathrm{rest}}=161\mu\mathrm{m}}$ for the cold component (fourth column of Table~\ref{tab:tab3}); this is a conservative choice since it includes the widest range of possible model uncertainties. 
Our fiducial estimates are therefore $T_{\mathrm{dust}}=87.1_{-18.3}^{+34.1}$, $\beta_{\mathrm{dust}}=0.91_{-0.49}^{+0.49}$, and $T_{\mathrm{dust}}=52.6_{-11.0}^{+10.3}$ K, $\beta_{\mathrm{dust}}=2.56_{-0.30}^{+0.37}$ for the warm and cold components, respectively. The former is consistent with expectations for small dust primarily radiating in the MIR \citep{Da_Cunha2008-jt}, while the latter is consistent with the typical temperatures and dust opacity indices of high-redshift starburst galaxies \citep{Magnelli2012-rf}, bolstering our interpretation of these two components as primarily AGN-heated and primarily star formation-heated.

\subsection{AGN-subtracted dust properties}

\begin{figure*}
\centering
\includegraphics[width=\textwidth]{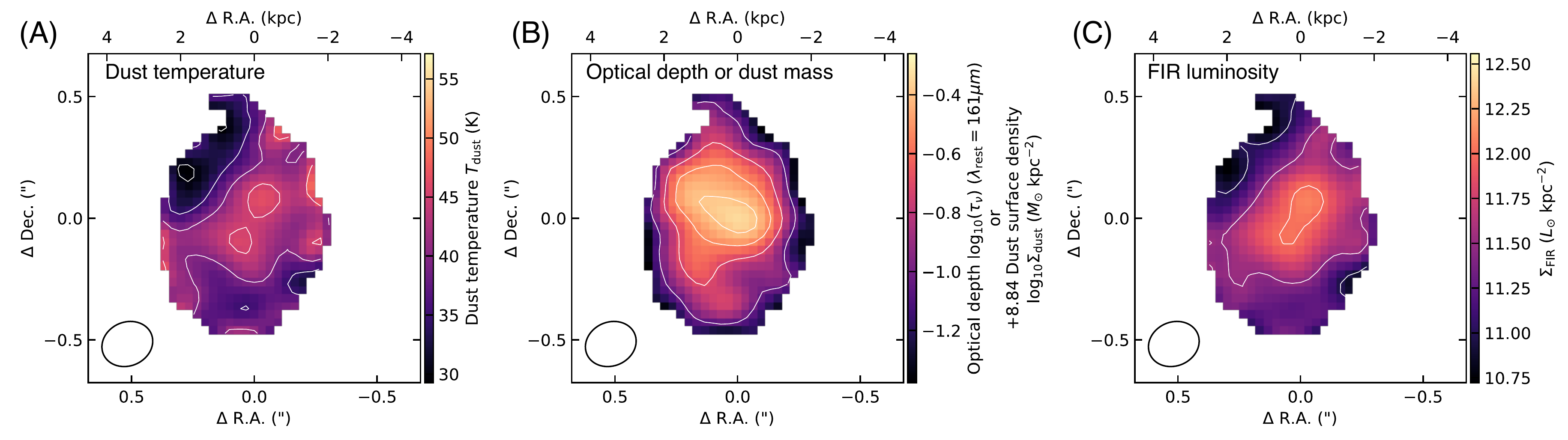}
\caption{The best-fit dust temperature (A), optical depth at Band 7 continuum $\lambda_{\mathrm{rest}}=161~\mu$m or dust mass surface density (B), and surface density of the integrated total IR luminosity over 8-1000$\mu$m, $\Sigma_{\mathrm{TIR}}$($L_\odot$ kpc$^{-2}$) (C). These are derived by fitting a greybody function at each pixel of the dust continuum images $I_{\lambda_{\mathrm{rest}}=90\mu\mathrm{m}}$ and $I_{\lambda_{\mathrm{rest}}=161\mu\mathrm{m}}$ after subtracting the point source (AGN-heated dust) contribution. The colour scales and contour levels are the same as in Fig.~\ref{fig:fig3} in order to aid in comparison. The size of the synthesized beam (FWHM) is shown in the lower-left corner of (A-C).\label{fig:fig12}}
\end{figure*}

Armed with our successful decomposition of the emission into a point-like AGN component and an extended star formation component, we are now also in a position to construct a map of the spatially-resolved dust properties in the galaxy with the AGN contamination removed. To this end, we start from the rest-frame $161~\mu\mathrm{m}$ and $90~\mu\mathrm{m}$ continuum images, and in each pixel, we subtract our best-fitting model estimate of the contribution from the point-like AGN-heated dust. We then repeat the fitting procedure presented in Section \ref{sec:result} on the point source-subtracted images. This procedure yields a map of properties for the cold dust component alone, rather than the unknown admixture of cold and hot components we obtain in Section \ref{sec:result}.
Figures~\ref{fig:fig12}~(A-C) show the temperature map, dust optical depth, and surface density of the SFR, respectively, derived in this manner. We again use Monte Carlo simulations to propagate the uncertainties, including not only the noise in the image and the uncertainty in the fiducial dust emissivity index $\beta=2.14\pm0.17$ as in Section \ref{sec:result}, but also our uncertainties on the properties of the point source component (see Table~\ref{tab:tab2}). 

\begin{figure}
\includegraphics[width=\columnwidth]{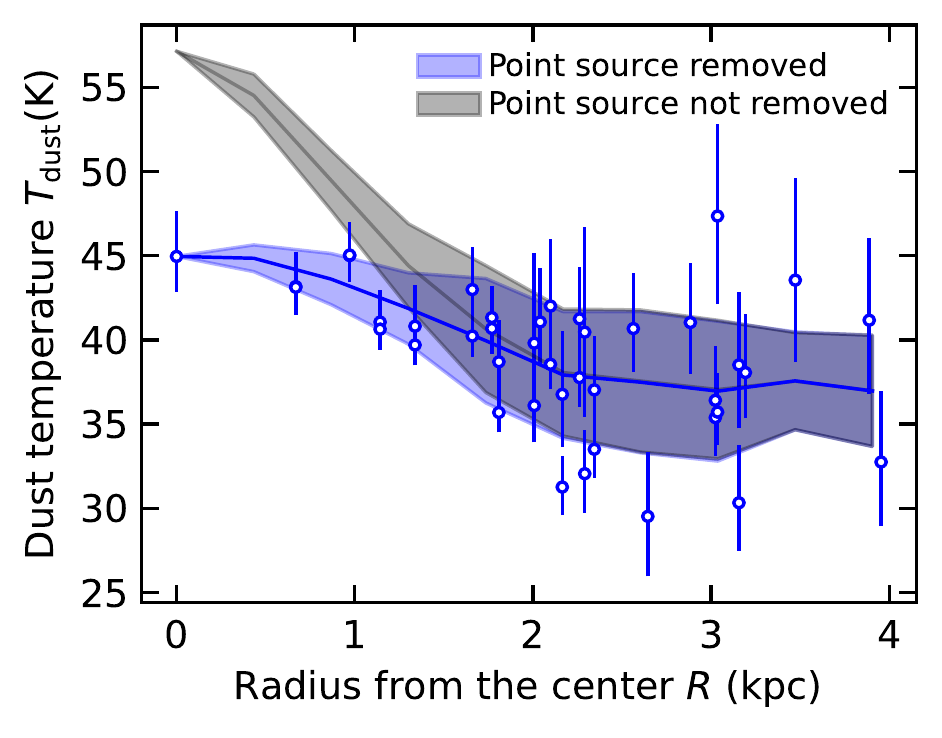}
\caption{Radial profile of the temperature of BRI 1335-0417, derived after subtracting the point source component (AGN-heated dust) shown in blue points and blue shade extracted from Fig.~\ref{fig:fig12}A in the same way as Fig.~\ref{fig:fig4}. The grey shade shows the radial profile derived before subtracting the point source component (same as the grey shade Fig.~\ref{fig:fig4}).
\label{fig:fig13}}
\end{figure}

Figure~\ref{fig:fig13} shows the one-dimensional dust temperature profile of the cold dust and the star formation density estimated after the warm dust component is removed. The temperature still shows a slight increase toward the centre, but does not show the steep temperature gradient seen in the results without the subtraction. The flat dust temperature indicates that the steep gradient in Fig.~\ref{fig:fig4} is entirely due to the warm dust heated by the AGN, suggested by the compact nature and the two-component dust SED fitting (Figs.~\ref{fig:fig10} and \ref{fig:fig11}), providing further confidence for the successful subtraction of the warm dust component heated by the central AGN. 

\begin{figure}
\includegraphics[width=\columnwidth]{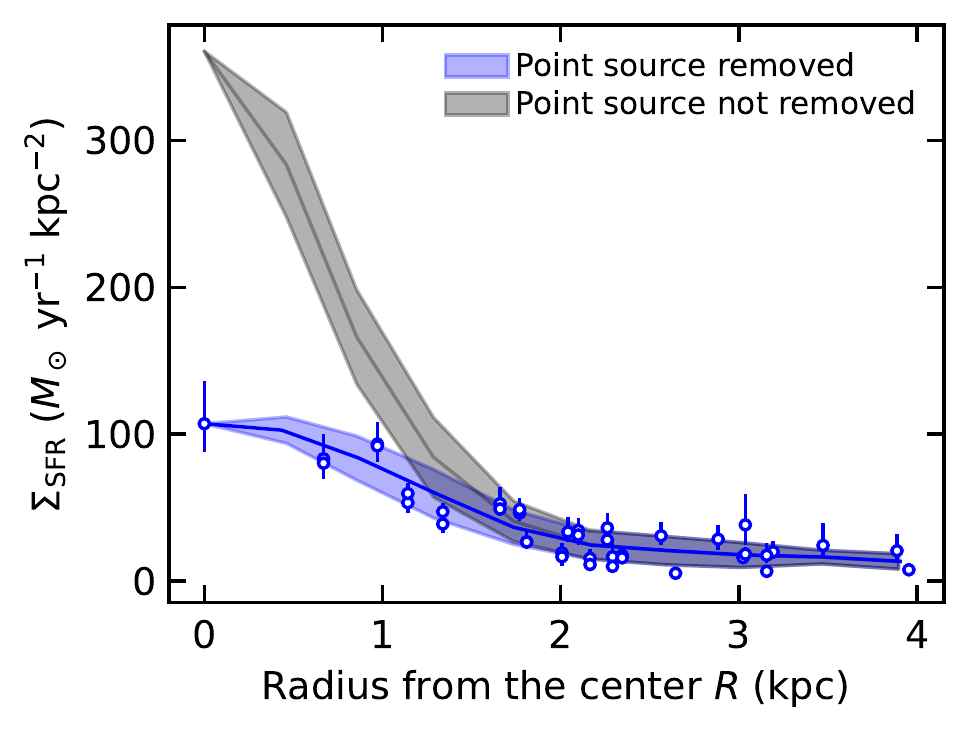}
\caption{Radial profile of star formation surface density $\Sigma_{\mathrm{SFR}}$ of BRI 1335-0417, derived after subtracting the point source component (AGN-heated dust) shown in blue points and blue shade. The points and shades are extracted from Fig.~\ref{fig:fig12}C in the same way as Fig.~\ref{fig:fig13} after converting the surface TIR luminosity density to SFR density (see text). The grey shade shows the radial profile of $\Sigma_{\mathrm{SFR}}$ derived without subtracting the point source component, derived by converting $\Sigma_\mathrm{TIR}$ as shown in Fig.\ref{fig:fig3}C to $\Sigma_\mathrm{SFR}$. \label{fig:fig14}}
\end{figure}

Figure~\ref{fig:fig14} shows a comparison of the SFR surface density $\Sigma_{\mathrm{SFR}}$ estimated from the FIR luminosity derived with and without removing the warm dust component heated by AGN (Figs. \ref{fig:fig3}C and \ref{fig:fig12}C). $\Sigma_{\mathrm{SFR}}$ is derived by $(\Sigma_{\mathrm{TIR}}/L_{\odot}\mathrm{kpc}^{-2})\times10^{-10} M_{\odot}\mathrm{kpc}^{-2}\mathrm{yr}^{-1}$ assuming the \citet{Chabrier2003-el} initial mass function, where $\Sigma_{\mathrm{TIR}}$ is measured for the cold dust component after subtracting the unresolved warm dust component from the images (Fig.\ref{fig:fig12}C). We see that if one does not remove the component, the central SFR surface density can be overestimated by over a factor of 3 at currently-available resolutions, and that the point source AGN-heated dust contaminates the derived dust physical properties out to $\sim$1 synthesized beam FWHM in radius ($\sim$ 1.3 kpc). In future higher-resolution observations, the contaminated region would presumably be smaller, but the contribution of the AGN-heated warm dust would be correspondingly larger in the central resolution elements, leading to an even larger overestimation of the central SFR surface density.

\begin{figure}
\includegraphics[width=\columnwidth]{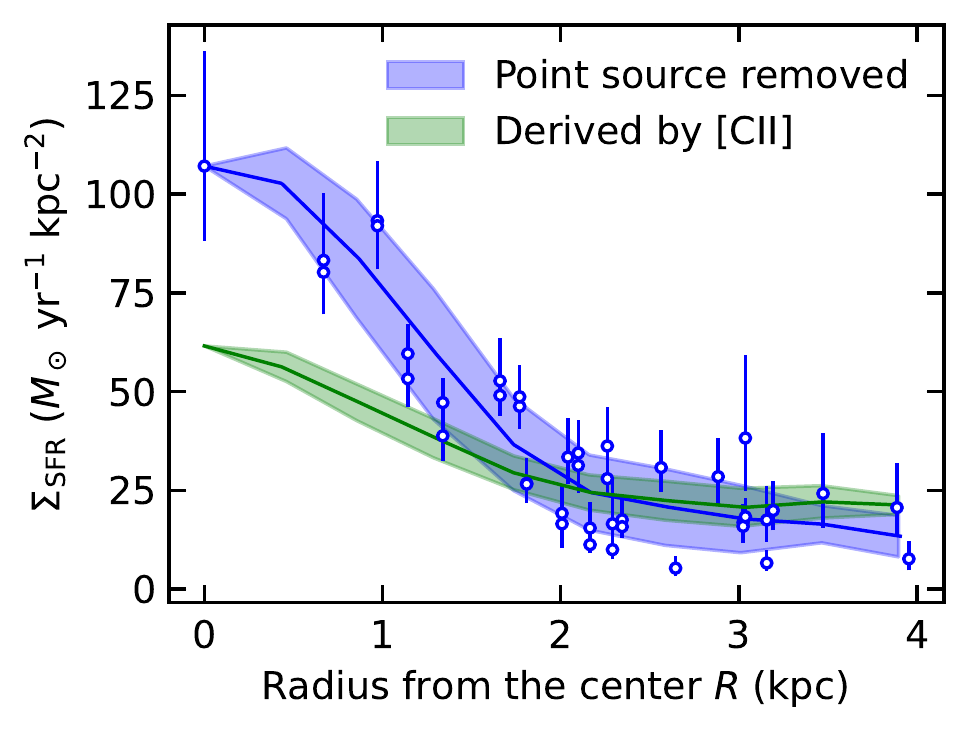}
\caption{Radial profile of star formation surface density $\Sigma_{\mathrm{SFR}}$ of BRI 1335-0417 derived from $L_{\mathrm{TIR}}$ shown in Fig.~\ref{fig:fig12}C, compared with the star formation surface density derived from [C~\textsc{ii}] line luminosity $L_{\mathrm{[C~\textsc{ii}]}}$ using the calibration of \citealt{De_Looze2011-jm}. \label{fig:fig15}}
\end{figure}

Figure~\ref{fig:fig15} compares the SFR surface density estimated from the point source-subtracted FIR luminosity with that estimated from the [C~\textsc{ii}] luminosity using the calibration of \citet{De_Looze2011-jm}. As expected from the central decrease in [C~\textsc{ii}]/$L_{\mathrm{TIR}}$, the [C~\textsc{ii}] luminosity under-predicts the SFR surface density in the central region. In the outer parts of the galaxy, the predicted SFR density from the [C~\textsc{ii}] line luminosity agrees well with that derived from $L_{\mathrm{TIR}}$. As a consistency check, we also compute the total SFR from the resolved map by integrating over it, using SFRs derived from [C~\textsc{ii}] in pixels where dust-based estimates are unavailable because the 68\% confident interval on $T_\mathrm{dust}$ is $> 10$ K (white regions in the image). The resulting integrated SFR is $\sim1476 M_{\odot}$ yr$^{-1}$, consistent with the SED modelling result $1700_{-400}^{+500} M_\odot$ yr$^{-1}$.

\section{Discussion}\label{sec:discussion}

The decomposition of BRI 1335-0417's emission into AGN- and star formation-powered components has implications both for the interpretation of the source itself and future studies of similar systems. We explore these in turn.

\subsection{Implications for the dynamics and history of BRI 1335-0417}

Our analysis allows us to draw a number of conclusions about the structure formation of BRI 1335-0417. First, our results suggest that the galaxy has an extended disk and a compact bulge both actively forming stars. Part of the evidence is dynamical: \citet{tsukui2021-vl} conclude based on [C~\textsc{ii}] kinematics that a compact mass $<1.3$ kpc in size resides in the centre of the galaxy. Our work here adds morphological evidence on top of this: both the $I_{\lambda_{\mathrm{rest}}=161\mu\mathrm{m}}$ and $I_{\lambda_{\mathrm{rest}}=90\mu\mathrm{m}}$ images require a compact Gaussian component with an effective radius $R_e$ of $\sim0.9$~kpc (calculated as $0.5\times$FWHM from the sizes obtained in Table~\ref{tab:tab2}) in addition to a nearly exponential [C~\textsc{ii}] disk template. The most natural explanation for the combined dynamical and morphological evidence is the presence of a compact, massive nuclear region that is actively star-forming. However, the fact that the rest-frame UV to optical part of the SED is best fit by a QSO template without dust extinction also suggests that the AGN of BRI 1335-0417 has cleared dust along our line of sight to the central accretion disk \citep[e.g.,][]{Fujimoto2022-li}, so gas is actively being ejected, and the central region may be in the last throes of star formation. Thus we conclude that BRI 1335-0417 may be in the transition from a QSO phase to being a passive bulge-dominated galaxy \citep{Hopkins2008-zh, Hopkins2008-fq}. With the compact star formation and the extended disk star formation, the galaxy may evolve into the passive bulged-disk system \citep{Fudamoto2022-tz} unless the galaxy is affected by external perturbation such as mergers.

\begin{figure*}
\centering
\includegraphics[width=\textwidth]{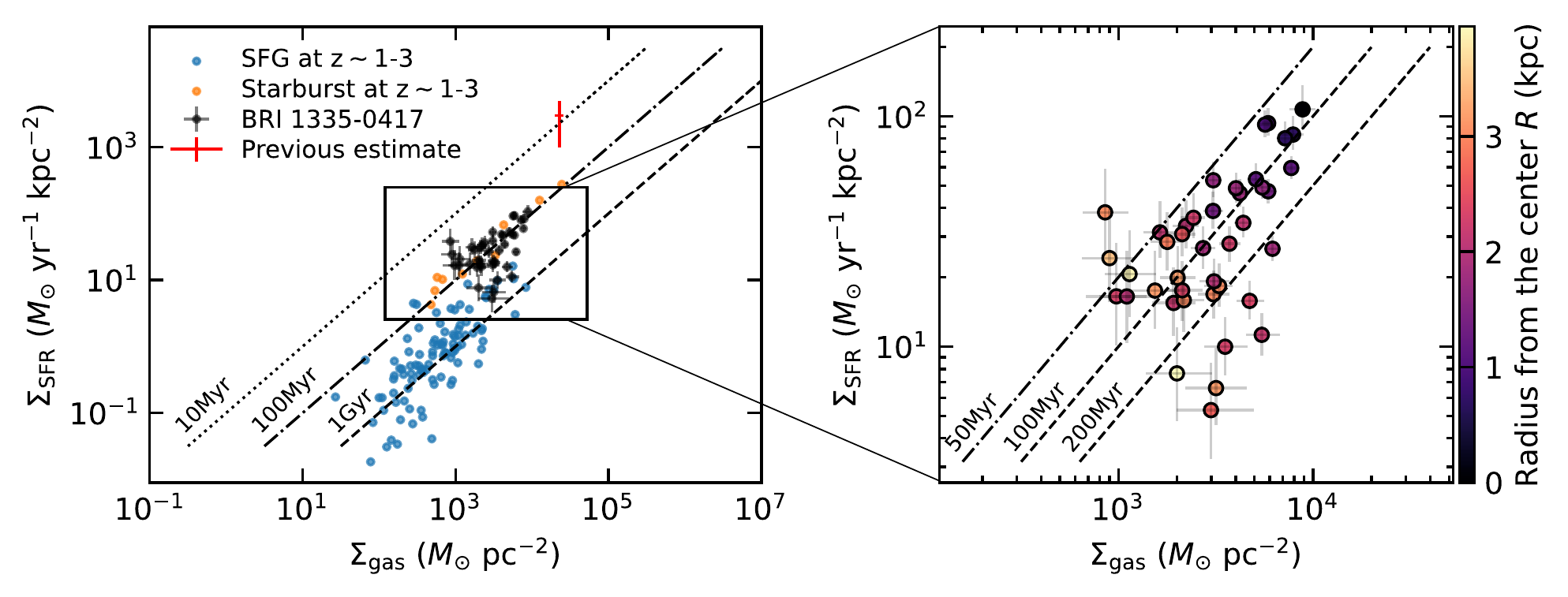}
\caption{The star formation rate surface density $\Sigma_{\mathrm{SFR}}$ and the gas surface density $\Sigma_{\mathrm{gas}}$ relation (Kennicutt–Schmidt diagram). Left: The black points indicate our spatially resolved measurements of BRI 1335-0417, where 3 pixels are extracted per beam, compared with the star-forming main sequence at redshift of 1-3 \citep{Tacconi2013-vp} and starburst galaxies at redshift of 1-3 \citep{Genzel2010-yp}. The red cross shows the previous estimate for BRI 1335-0417 \citep{Jones2016-cj}, refined by our new spatially resolved SFR estimate with the AGN-heated dust component removed. Right: Same as the black points in the left, but colour-coded with the radial distance from the centre. \add{The diagonal lines in both panels indicate linear relations between $\Sigma_{\mathrm{SFR}}$ and $\Sigma_{\mathrm{gas}}$ corresponding different gas depletion timescale $\Sigma_{\mathrm{gas}}/\Sigma_{\mathrm{SFR}}$.} \label{fig:fig16}}
\end{figure*}

Our results also explain the organized disk rotation in the galaxy. In Figure~\ref{fig:fig16}, we show pixel-by-pixel measurements of $\Sigma_{\mathrm{gas}}$ vs.~$\Sigma_{\mathrm{SFR}}$ (i.e., the Kennicutt-Schmidt diagram) for BRI 1335-0417. The gas mass is derived by converting the spatially resolved dust mass surface density in Fig.~\ref{fig:fig3}B to a total mass surface density assuming the same gas-to-dust ratio of $54.2\pm9.3$ estimated in Sec.~\ref{sec:result}. We find that most pixels are located in the starburst regime characterized by gas depletion times of 50-200 Myr \citep{Genzel2010-yp}, much less than the $\sim 1$ Gyr values typically found for main sequence star-forming galaxies at redshift of 1-3 \citep{Tacconi2013-vp}. If the starburst is triggered by a gas-rich major merger, which violently disturbs the gas kinematics, it may take at least one orbital period for the gas to settle into organized disk rotation. The 50-200 Myr depletion time is comparable to the $\approx 120$ Myr orbital period at the disk effective radius derived from [C~\textsc{ii}] kinematics \citep{tsukui2021-vl}. This explains why the galaxy has had time to settle into relaxed, disk-like kinematics, at least in the centre where the orbital period is shortest; the outer disk, where the orbital time is longer, is also rotating, but shows significant morphological disturbance \citep{tsukui2021-vl}, consistent with a system age near the upper end of the depletion time range.

It is worth noting that our estimated gas depletion times are much larger than the previous estimates by \citet{Jones2016-cj}, who not only found a much higher SFR due to AGN contribution, but also adopted a size of 1.8 kpc$^2$ for the area of the star-forming disk based on the 44 GHz continuum, much smaller than the size revealed in our spatially resolved maps. Both factors contribute to an underestimate of the depletion time. Using this underestimate leads to the conclusion that the system must be less than one dynamical time old, which is hard to reconcile with the kinematics as noted in Sec.~\ref{sec:intro}. Our increased depletion time resolves this discrepancy.

Finally, we note that Fig.~\ref{fig:fig16} also shows that the central region of the galaxy has shorter gas depletion time on average than the disk (outer) part of the galaxy, while the outer part has a larger scatter in depletion time. The large scatter reflects the high/low-temperature regions aligned with the minor/major axis (Fig.~\ref{fig:fig6}), and suggests that the SFR in part of the outer disk may be locally enhanced by cold gas/satellite accretion from a preferential direction, or by the AGN wind to the perpendicular direction to the disk as discussed in Sec.~\ref{subsec:subsec35}. This combined with the shorter depletion time in the centre suggests that the galaxy may quench star formation inside out to form a bulge-disk system \citep{Van_Dokkum2015-sf,Tacchella2015-xa} due to the combination of the removal of the gas by the AGN wind, earlier consumption of the gas in the centre, and/or gas supply from ongoing gas/satellite accretion supporting the star formation in the disk.

\subsection{Implications for future studies of high-redshift starbursts and QSO hosts}

In BRI 1335-0417, the warm dust component heated by AGN contributes only a small fraction of the total observed flux, $\sim$15\% and $\sim$21\% at $\lambda_{\mathrm{rest}}=161\mu$m and $\lambda_{\mathrm{rest}}=90\mu$m, respectively. However, ignoring this component leads to a factor of three overestimate of the total SFR (see Fig.~\ref{fig:fig2} and Fig.~\ref{fig:fig10}), because the warm dust component has high temperature, $T_\mathrm{dust}=87.1^{+34.1}_{-18.3}$, and thus contributes a substantially larger fraction of the inferred total TIR luminosity ($L_{\mathrm{TIR}}\propto T_{\mathrm{dust}}^{4+\beta}$). The warm dust component also contributes a significant fraction of the flux within the central resolution element, leading to a factor of $\sim 4$ overestimate of the local SFR density  (see Fig.~\ref{fig:fig14}), and contaminates the SFR density estimate out to roughly the FWHM of the beam in radius. Thus the point source subtraction is crucial for accurate estimates of both the total SFR and its spatial distribution, \textit{even when the AGN contribution to the observed bands is relatively small}.

Nor does higher resolution remove this need. In a higher-resolution observation, the region contaminated by the AGN is smaller, but the AGN contribution is correspondingly greater because the relative fraction of the extended cold dust component becomes smaller in the central beam. As an extreme example of this, \citet{Walter2022-uh} reported 200 pc resolution observation towards the quasar J234833.34–305410.0 at redshift 6.9 and estimated the dust temperature of the central resolution element (radius of 110pc) to be $>$132K, and the star formation rate density to be $25500M_\odot$yr$^{-1}$kpc$^{-2}$ with the assumption that the AGN is negligible. The effective radius of the resolution element 110 pc is much smaller than our study ($r=720$ pc). As the authors discussed, the compact emission from the AGN-heated dust may dominate in the central resolution element. Our findings here strongly support that conjecture.

In this paper, we decompose the AGN and star formation contributions using ALMA observations of the FIR part of SED, where the ratio of AGN to galactic luminosity is less extreme than at rest-frame UV to optical, where the black hole accretion disk greatly outshines the host galaxy \citep{Marshall2020-zf}. Recently \citet{Ding2022-nd} demonstrated that James Web Space Telescope (JWST) data in rest-frame optical also enable the decomposition of the surface brightness distribution into the point source AGN component and extended host galaxy stellar component. The separation of AGN-host galaxies with JWST, coupled with decomposition in the FIR and SED modelling such as that we have demonstrated here, will provide measurements of the already-formed stellar masses in addition to the SFR. These complementary measurements will provide the information crucial to understanding how the galaxy-BH relation is set in the early universe. 

\section{Conclusion and summary}\label{sec:conclusion}
We present $\sim$1kpc ALMA imaging of the dust continuum, [C~\textsc{ii}] emission, and CO(7-6) emission of a $z=4.4$ quasar host galaxy, BRI1335-0417. Due to the unique brightness of the galaxy, the observations provide a number of resolution elements across the galaxy, allowing us to study the spatially resolved ISM properties of the host galaxy, and to disentangle the light coming from the unresolved, AGN-powered region from that produced by star formation in the surrounding galaxy. 
Our main findings and their implications are the following.

\begin{itemize}
    \item Using the spatially resolved continuum images $I_{\lambda_{\mathrm{rest}}=161\mu\mathrm{m}}$ and $I_{\lambda_{\mathrm{rest}}=90\mu\mathrm{m}}$, but without first separating the AGN and galaxy components, we constrain the shape of the greybody for individual pixels and derive the dust temperature, optical thickness, and dust mass surface density. The central resolution element is found to be optically thick $\tau_{\nu}\sim 0.6$ at $\lambda_{\mathrm{rest}}=161\micron$ and $\tau_{\nu}\sim 1.3$ at $\lambda_{\mathrm{rest}}=90\micron$. The temperature shows a steep radial gradient towards the centre, reaching $57.7 \pm 0.4$ K, which is higher than the typical temperatures of 47K and 40K for quasar host galaxies \add{\citep{Beelen2006-fc}} and star-forming galaxies \add{\citep{Magnelli2012-rf}}, respectively. 

    \item The pixel-by-pixel temperature distribution image shows that high-temperature peaks are preferentially aligned with the disk minor axis. The anisotropic distribution coincides with the high velocity dispersion region of [C~\textsc{ii}], both of which show a conical shape aligned with the disk minor axis. With current data, we cannot conclude whether the feature is due to the AGN wind heating the dust \citep{Saito2017-mb} or to the presence of a star-forming region driven by anisotropic cold gas supply from gas/satellite accretion \citep{Dekel2009-fa}. 
        
    \item Image decomposition analysis reveals the presence of a point source in the two dust continuum images $I_{\lambda_{\mathrm{rest}}=161\mu\mathrm{m}}$ and $I_{\lambda_{\mathrm{rest}}=90\mu\mathrm{m}}$, whose positions coincide with the highest temperature, the peak of the dust continuum image, and the optical quasar position. The point source contribution to the total flux is small, $\sim$15\% and $\sim$21\% for $I_{\lambda_{\mathrm{rest}}=161\mu\mathrm{m}}$ and $I_{\lambda_{\mathrm{rest}}=90\mu\mathrm{m}}$, respectively. However, in the central resolution element, the contribution is much larger, 49.8$_{-1.3}^{+1.5}$\% and 56.7$^{-6.1}_{+4.5}$\% for the respective images to the flux. 
    
    \item We model the FIR SED assuming that the point-source flux comes from a warm dust component heated by the AGN, and the remainder comes from cold dust component heated by the star-forming host galaxy. The decomposed fluxes constrain the AGN-heated warm dust contribution to the FIR part of the SED, which is usually assumed to be dominated by the cold dust heated by star formation. We estimate temperatures of $T_\mathrm{dust}=87.1^{34.1}_{-18.3}$ K and $T_{\mathrm{dust}}=52.6^{+10.3}_{-11.0}$ K for the warm and cold components, respectively. We estimate the SFR from the FIR luminosity of the cold component, finding a SFR of $1700_{-400}^{+500} M_\odot$ yr$^{-1}$. This is a factor of three less than the previously estimated value $5040\pm{1300} M_\odot$ yr$^{-1}$ due to the high AGN fraction in the FIR luminosity $53^{+14}_{-15}$\%.
    
    \item The SED fit suggests that there are two dust components with different temperatures in the central resolution element in the images. The central temperature of $57\pm 0.3$ K we obtain without decomposing the AGN and star formation components should be interpreted as the luminosity-averaged solution within the beam, which results from fitting these two dust components with a single greybody function. After removing the point source (warm dust component) component from the images, we remeasure the dust properties for the star formation-heated cold dust component only. This fit shows a nearly flat temperature profile with a slight increase toward the centre, suggesting that the steep temperature gradient found in the one-component fit is entirely due to the unresolved warm dust component. The single-component fit overestimates the central SFR density by a factor of \add{$\sim$}4 as a result.
    
    \item Our estimates of star formation surface density $\Sigma_{\mathrm{SFR}}$ after AGN subtraction and gas surface density $\Sigma_{\mathrm{gas}}$ for individual pixels show a roughly linear sequence with a gas depletion time of 50-200Myr. This places the galaxy in the starburst regime, clearly separated from main-sequence galaxies at $z\sim1-3$ with gas depletion times of $\sim$1 Gyr, but makes it a typical starburst rather than an extreme outlier.
    
\end{itemize}

Through this study, we demonstrate a method to constrain the SED shape of an unresolved warm dust component heated by AGN and an extended cold dust component in the host galaxy by combining spatially resolved information in two images $I_{\lambda_{\mathrm{rest}}=161\mu\mathrm{m}}$ and $I_{\lambda_{\mathrm{rest}}=90\mu\mathrm{m}}$ with integrated UV to FIR SED analysis. This method provides the total spatially resolved SFR, with contamination from warm dust heated by the AGN removed allowing us to study how quasar activity affects stellar mass build-up. Our method and recently demonstrated AGN-host galaxy decomposition with JWST \citep{Ding2022-nd} provide complementary measurements crucial to understanding how the galaxy-BH relation is set in the early universe.

\section*{Acknowledgements}

TT is grateful for the helpful discussions with Satoru Iguchi, Takuya Hashimoto, Kentaro Nagamine, and Yuichi Matsuda. This research was supported by the Australian Research Council Centre of Excellence for All Sky Astrophysics in 3 Dimensions (ASTRO 3D), through project number CE170100013. 
MRK acknowledges support from the Australian Research Council through Laureate Fellowship FL220100020. Data analysis was carried out on the Multi-wavelength Data Analysis System operated by the Astronomy Data Center (ADC), National Astronomical Observatory of Japan. This paper makes use of the following ALMA data: ADS/JAO.ALMA\#2017.1.00394.S, and \#2018.1.01103.S. ALMA is a partnership of ESO (representing its member states), NSF (USA) and NINS (Japan), together with NRC (Canada), NSC and ASIAA (Taiwan) and KASI (Republic of Korea), in cooperation with the Republic of Chile. The Joint ALMA Observatory is operated by ESO, AUI/NRAO and NAOJ. \add{This research is based on observations made with the NASA/ESA Hubble Space Telescope obtained from the Space Telescope Science Institute, which is operated by the Association of Universities for Research in Astronomy, Inc., under NASA contract NAS 5–26555. These observations are associated with program(s) GO 8572.} This work has made use of data from the European Space Agency (ESA) mission
{\it Gaia} (\url{https://www.cosmos.esa.int/gaia}), processed by the {\it Gaia}
Data Processing and Analysis Consortium (DPAC,
\url{https://www.cosmos.esa.int/web/gaia/dpac/consortium}). Funding for the DPAC
has been provided by national institutions, in particular the institutions
participating in the {\it Gaia} Multilateral Agreement. This research has made use of the NASA/IPAC Infrared Science Archive, which is funded by the National Aeronautics and Space Administration and operated by the California Institute of Technology. This paper makes use of the Herschel, which is an ESA space observatory with science instruments provided by European-led Principal Investigator consortia and with important participation from NASA. \add{This work is based in part on archival data obtained with the Spitzer Space Telescope, which was operated by the Jet Propulsion Laboratory, California Institute of Technology under a contract with NASA. Support for this work was provided by an award issued by JPL/Caltech.} This research has made use of the NASA/IPAC Extragalactic Database (NED) which is operated by the Jet Propulsion Laboratory, California Institute of Technology, under contract with the National Aeronautics and Space Administration. 

\section*{Data Availability}
The ALMA data we use in this work \add{are} publicly available in \url{https://almascience.nrao.edu/aq/}. \add{The HST data we use in this work can be obtained in \url{https://mast.stsci.edu/portal/Mashup/Clients/Mast/Portal.html}.} \add{The photometric data from Spitzer and Herschel we use in this work are publicly available at doi: \href{https://www.ipac.caltech.edu/doi/irsa/10.26131/IRSA3}{10.26131/IRSA3} and \citet{Marton2017-nl}, respectively.}



\bibliographystyle{mnras}
\bibliography{example} 




\appendix

\counterwithin{figure}{section}
\counterwithin{table}{section}
\section{Supplemental figures}

\begin{figure}
\centering
\includegraphics[width=\columnwidth]{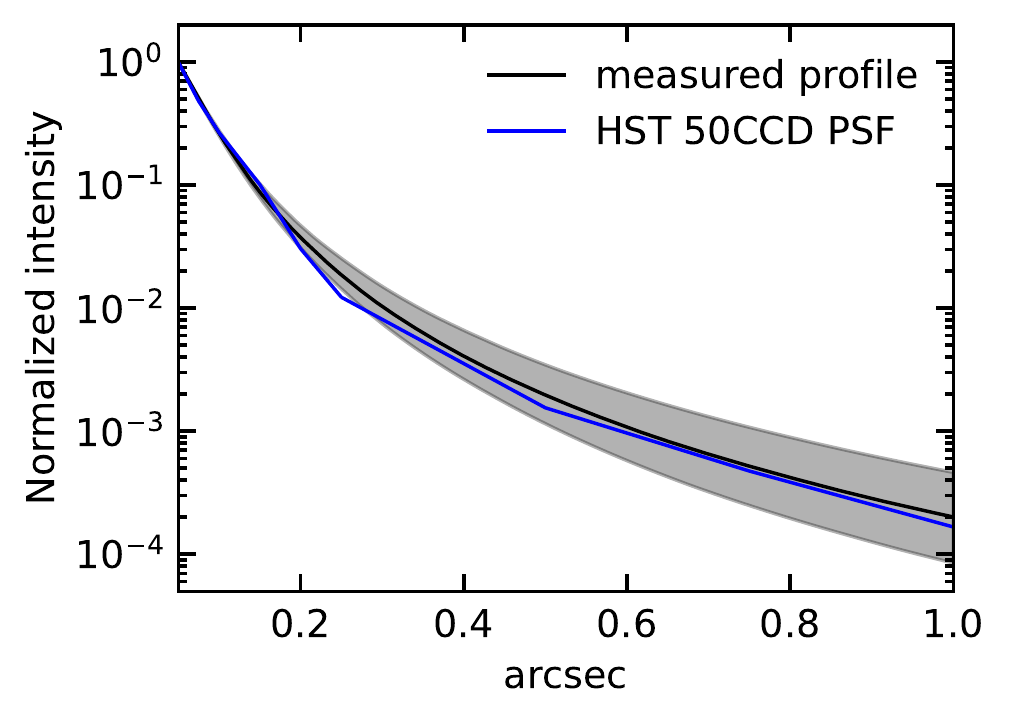}
\caption{\add{The radial distribution of \textit{HST} STIS/50CCD image (black line) and the point spread function \citep[blue line: ][]{Medallon_S_and_Welty_D_et_al_undated-wi}. The profile is measured by fitting a Moffat function to the image (Fig.~\ref{fig:fig0}). The grey shaded region is the $1\sigma$ uncertainty of the measured profile due to statistical noise. This comparison demonstrates that the optical emission in the STIS/50CCD image is consistent with a point source within the uncertainty. \label{fig:figA0}}}
\end{figure}

\begin{figure*}
\centering
\includegraphics[width=\textwidth]{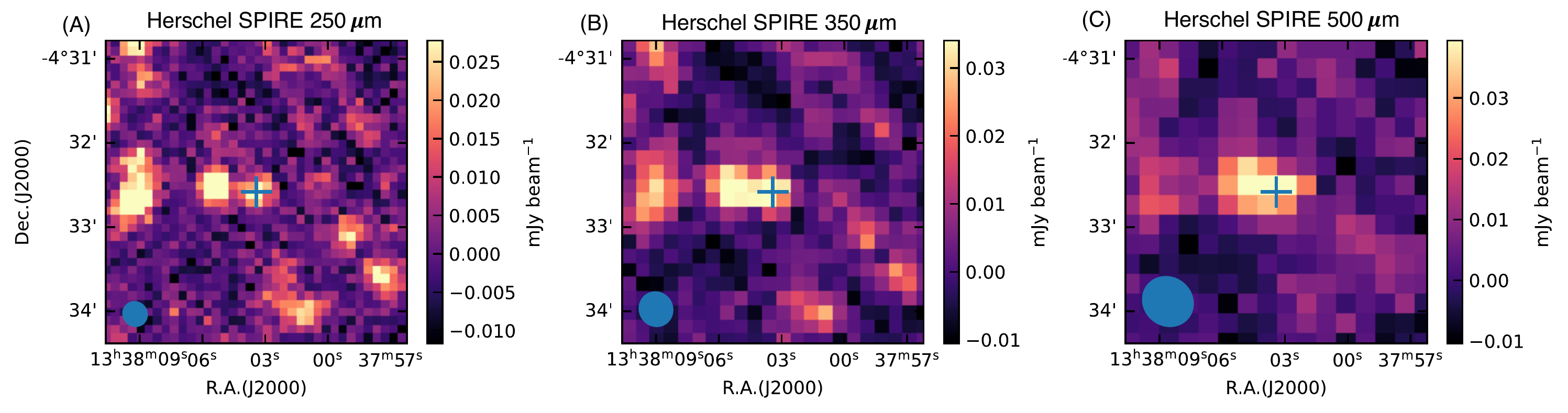}
\caption{Herschel SPIRE images at 250~$\micron$ (A), 350~$\micron$ (B), and 500~$\micron$ (C) bands for BRI 1335-0417. The blue cross shows the source position of BRI 1335-0417. The blue ellipse at the left corner of each panel is the size of the point spread function (FWHM). At 250~$\micron$ with the highest resolution, the galaxy is separated from the nearby bright source at East. At 350~$\micron$, 500~$\micron$ bands, the galaxy is not separated from and can be contaminated by the neighbouring bright source. Therefore, we did not include the fluxes of 350~$\micron$ and 500~$\micron$ for our SED modelling. \label{fig:figa1}}
\end{figure*}

\begin{figure*}
\centering
\includegraphics[height=0.27\textheight]{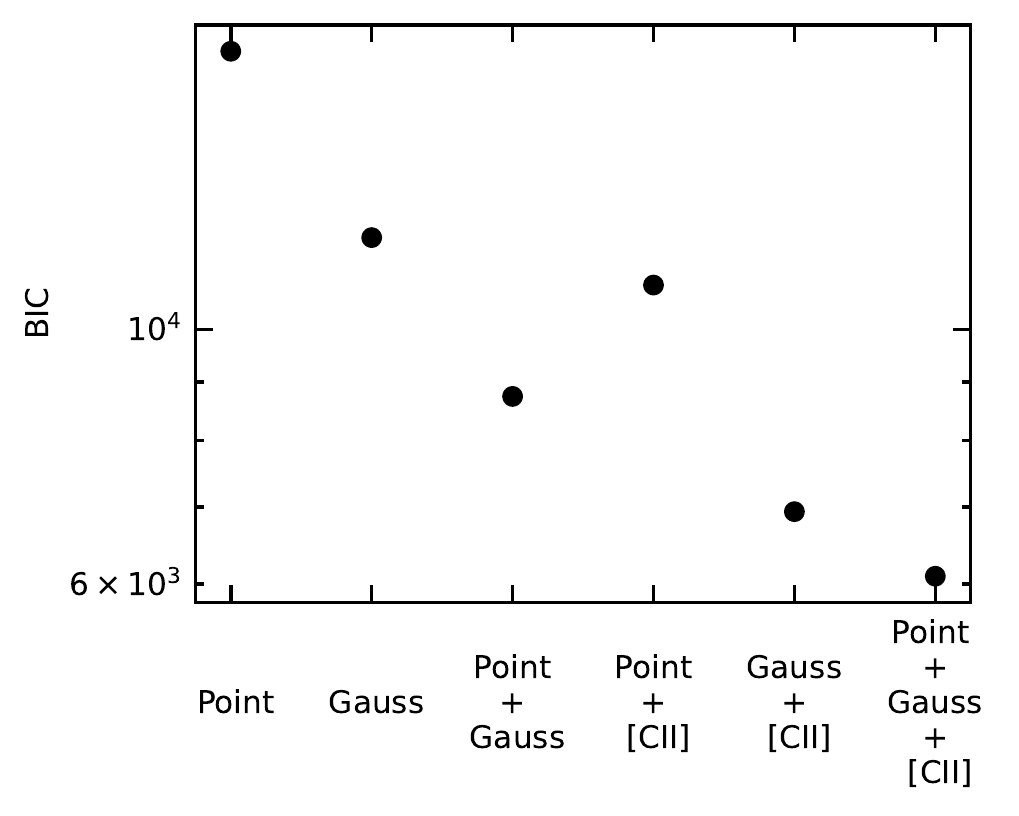}
\includegraphics[height=0.27\textheight]{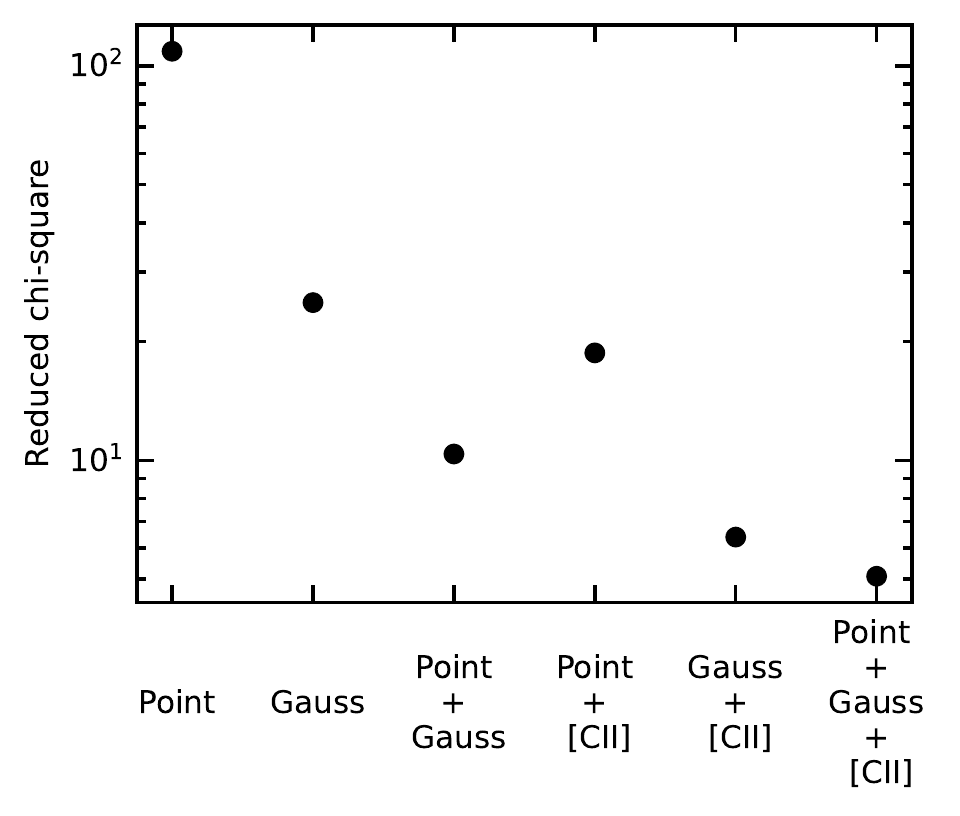}
\caption{The reduced chi-square values (left) and Bayesian information criterion (BIC: right) for all combinations of model components used to fit Band 7 continuum image, $I_{\lambda_{\mathrm{rest}}=161\mu\mathrm{m}}$. \label{fig:figa2}}
\end{figure*}

\begin{figure*}
\centering
\includegraphics[height=0.27\textheight]{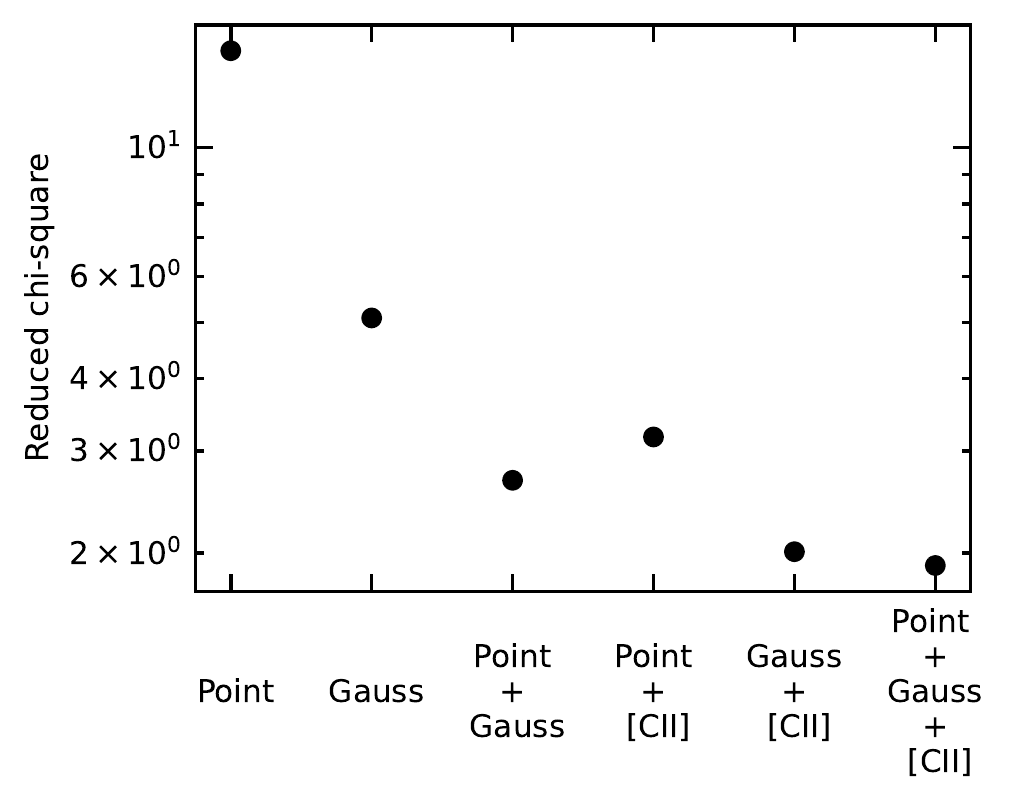}
\includegraphics[height=0.27\textheight]{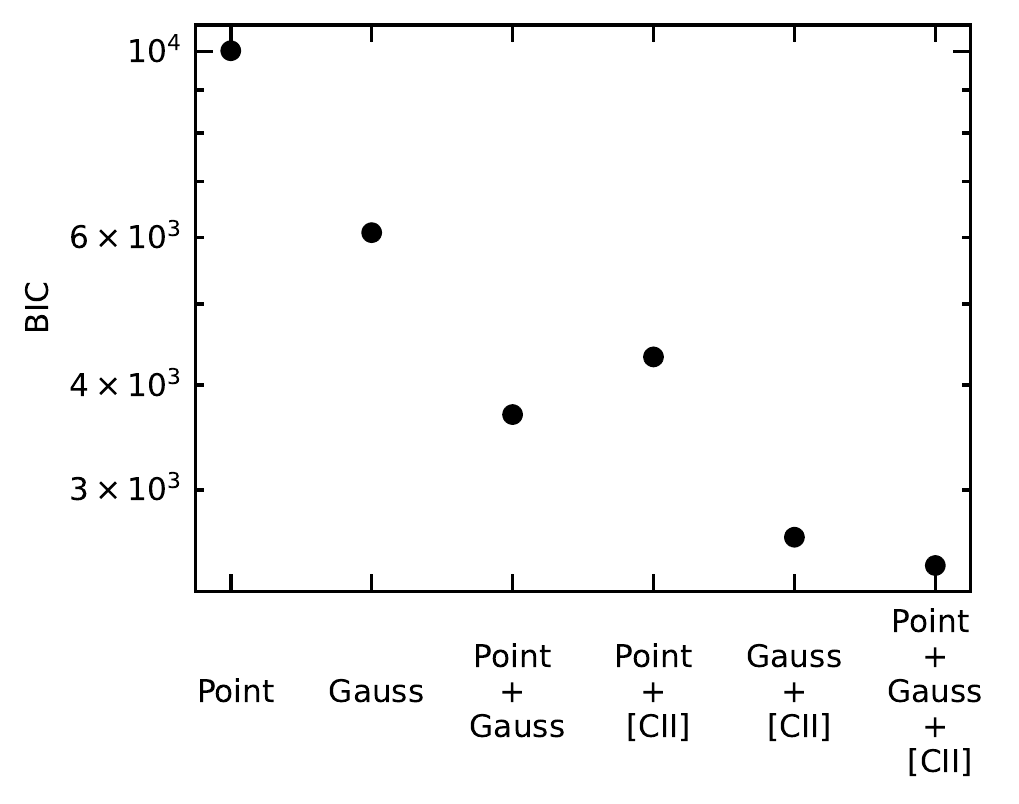}
\caption{Same as Fig.~\ref{fig:figa2} for Band 9 continuum image, $I_{\lambda_{\mathrm{rest}}=90\mu\mathrm{m}}$.\label{fig:figa3}}
\end{figure*}

\begin{figure*}
\centering
\includegraphics[width=0.96\textwidth]{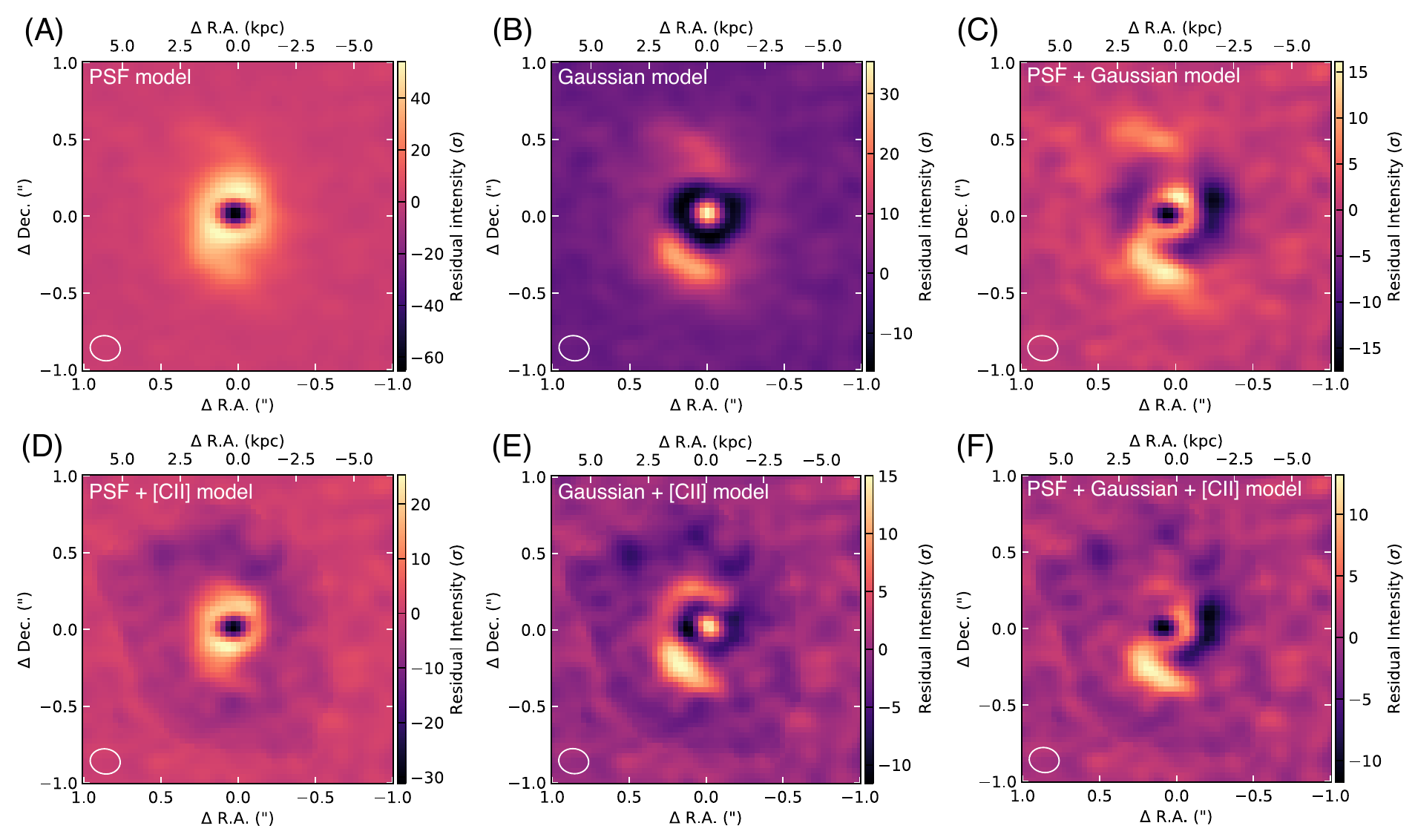}
\caption{The residual of observed data minus best-fit model for all combinations of model components used to fit the Band 7 continuum image, $I_{\lambda_{\mathrm{rest}}=161\mu\mathrm{m}}$. \label{fig:figa4}}
\end{figure*}

\begin{figure*}
\centering
\includegraphics[width=0.96\textwidth]{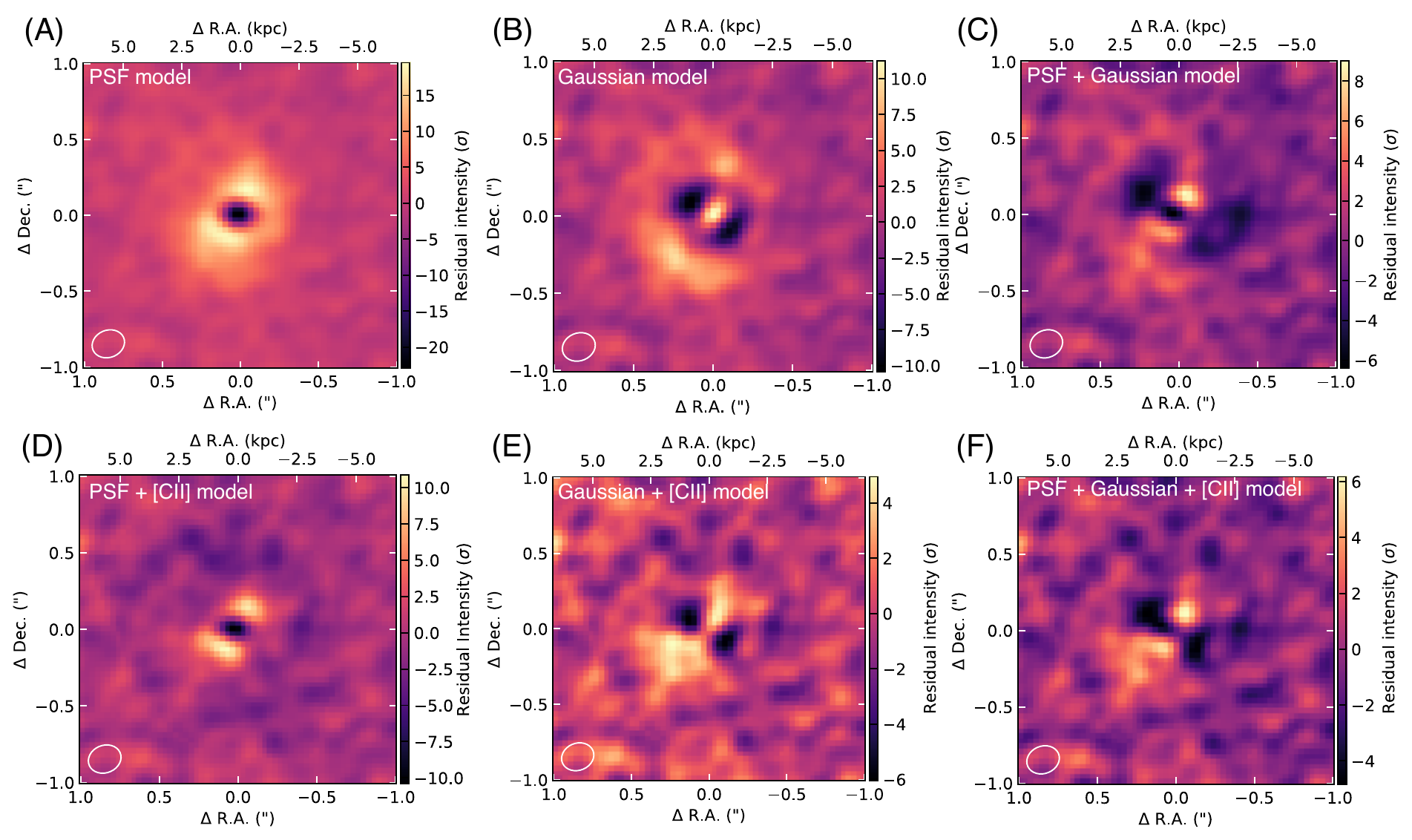}
\caption{The same as Fig.~\ref{fig:figa4} for the Band 9 continuum image, $I_{\lambda_{\mathrm{rest}}=90\mu\mathrm{m}}$. \label{fig:figa5}}
\end{figure*}

\begin{figure*}
\centering
\includegraphics[width=\textwidth]{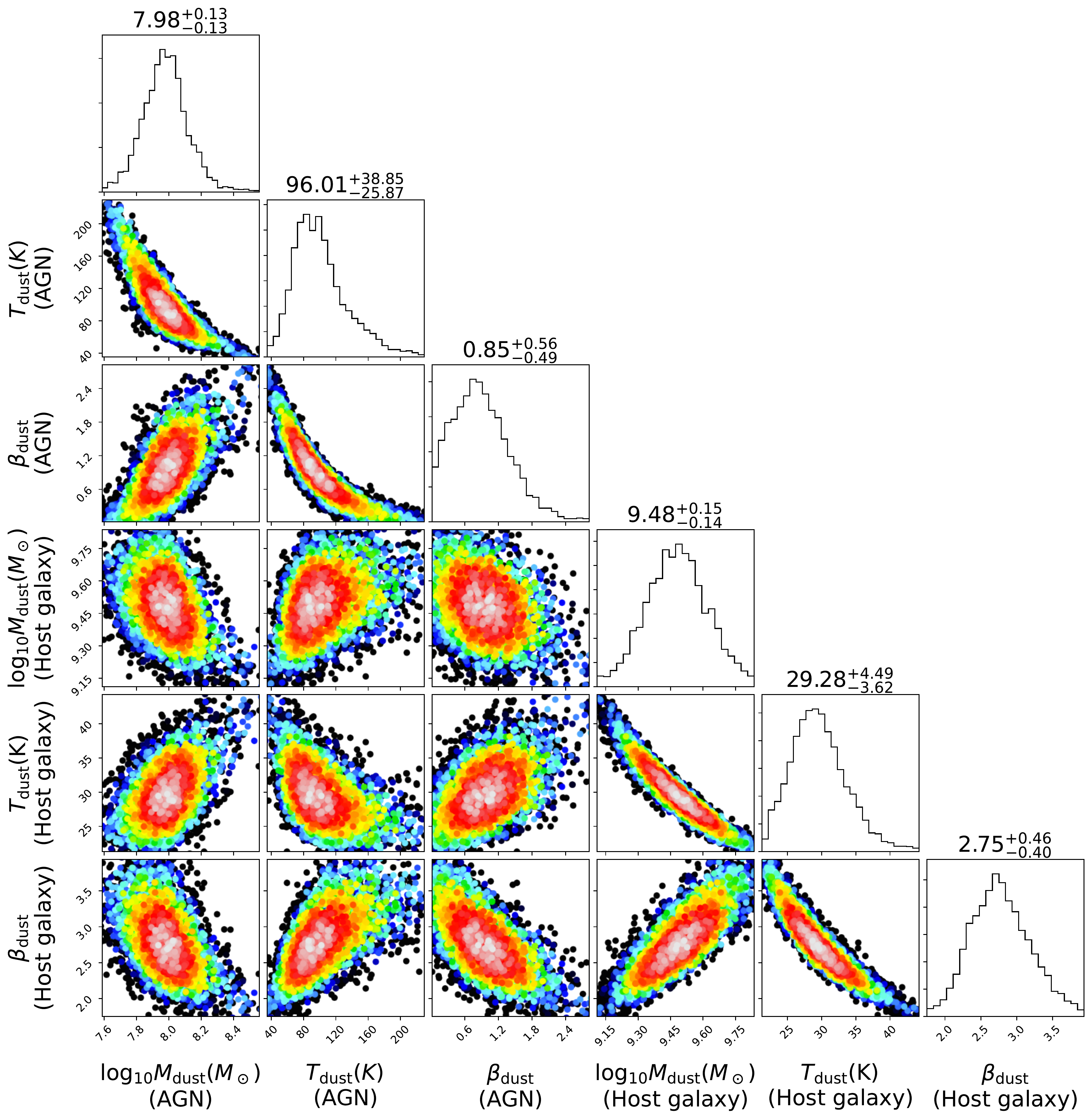}
\caption{Posterior probability distributions of the dust mass $M_{\mathrm{dust}}$, dust temperature $T_{\mathrm{dust}}$ and dust emissivity $\beta_{\mathrm{dust}}$ of two greybody functions fitted to FIR part of the SED, sampled using \textsc{emcee} code \citep{Foreman-Mackey2013-yn}. The colour in the sampled distribution indicates the relative log-likelihood of the sample; blue shows the least likely to white the most, while black shows the even less likely points with $\Delta \chi^{2}>9$, corresponding to 3$\sigma$ confidence interval in $\chi^{2}$ statistics. The derived parameters are summarized in Table~\ref{tab:tab3}. \label{fig:figa6}}
\end{figure*}

\begin{figure*}
\centering
\includegraphics[width=\textwidth]{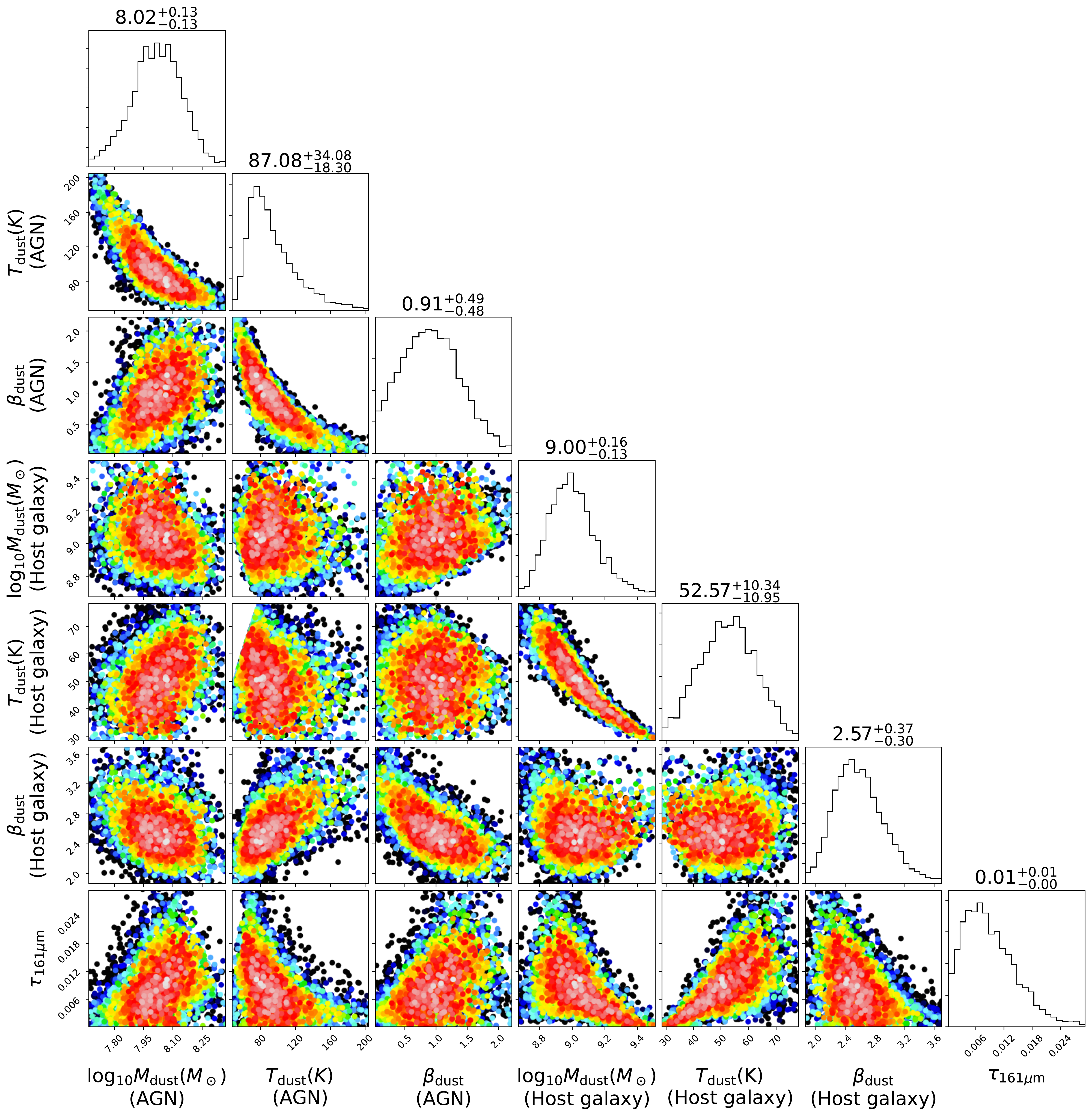}
\caption{Same as Fig \ref{fig:figa6} for two greybody functions with free optical depth at rest-frame 161~$\micron$, $\tau_{161\mu \mathrm{m}}$ for the cold dust component. We used prior on the temperature of the warm dust component does not exceed the cold dust component. The derived parameters are summarized in Table~\ref{tab:tab3}. \label{fig:figa7}}
\end{figure*}

\begin{figure*}
\centering
\includegraphics[width=0.6\textwidth]{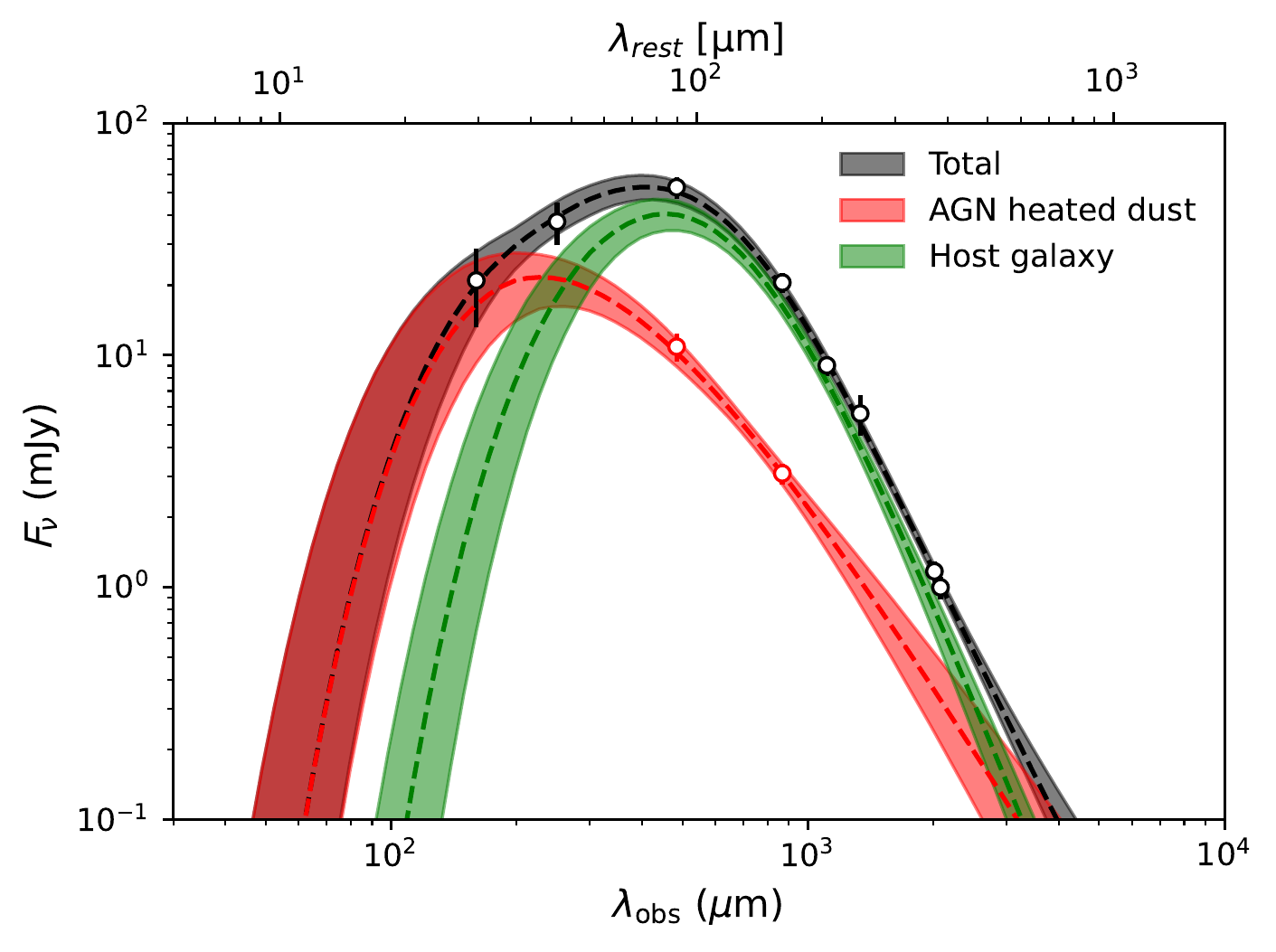}
\caption{Same as Fig \ref{fig:fig10}, but for the fit relaxing the optically thin assumption with the free optical depth at rest-frame 161~$\micron$, $\tau_{\lambda_{\mathrm{rest}}=161\mu \mathrm{m}}$. Optically thin case is included in the range of the confidence interval of the model fitting. The overall shape of the model does not change and thus provides consistent parameters such as SFR, and AGN fraction with the optically thin case (see Table~\ref{tab:tab3}). \label{fig:figa8}}
\end{figure*}

\begin{figure*}
\centering
\includegraphics[width=0.7\textwidth]{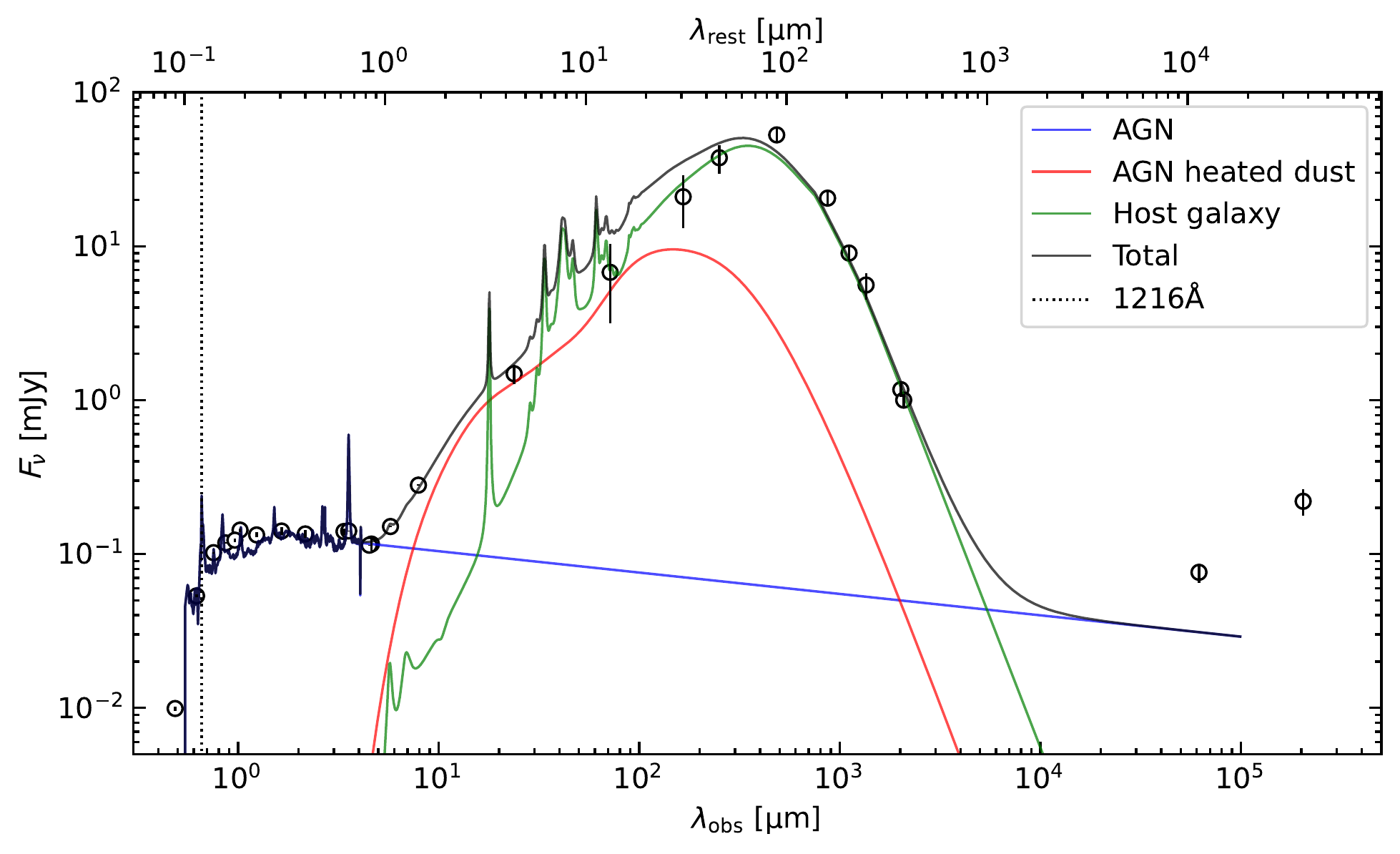}
\caption{The UV to Radio spectral energy distribution of BRI1335-0417 (black points, see Table~\ref{tab:taba1}) with the best-fit \textsc{stardust} model \citep{Kokorev2021-yp} without using point source fluxes (red points) at rest-frame $161~\mu\mathrm{m}$ and $90~\mu\mathrm{m}$ as AGN-heated dust component. The best-fit model returns the SFR estimate of $5260\pm31 M_\odot$ yr$^{-1}$ much higher than $1770\pm20 M_\odot$ yr$^{-1}$ estimated using point source fluxes at rest-frame $161\mu \mathrm{m}$ and $90\mu \mathrm{m}$ bands as emission from AGN-heated dust. \label{fig:figa9}}
\end{figure*}

\begin{table*}
\begin{tabular}{c c c c c}
\hline\hline
Instrument &  Wavelength (micron) & Flux(mJy) & Flux uncertainty (mJy) & Sources\\
\hline
Pan-STARRS1-g  &    0.477 &  0.0099 &  3.100e-04 & \citep{Chambers2016-ik}\\
Pan-STARRS1-r  &    0.612 &  0.053 &  1.300e-03 &\citep{Chambers2016-ik} \\
Pan-STARRS1-i  &    0.747 &  0.102 &  1.000e-03 &\citep{Chambers2016-ik}\\
Pan-STARRS1-z  &    0.865 &  0.118 &  1.000e-03 &\citep{Chambers2016-ik}\\
Pan-STARRS1-y  &    0.960 &  0.123 &  3.000e-03 &\citep{Chambers2016-ik}\\
VISTA-Y    &    1.018 &  0.143 &  4.000e-03 &\citep{Lasker2021-lr}\\
2MASSJ    &    1.239 &  0.133 &  5.000e-03 &\citep{Lasker2021-lr}\\
2MASSH    &    1.649 &  0.141 &  8.000e-03 &\citep{Lasker2021-lr}\\
2MASSKs   &    2.163 &  0.135 &  8.000e-03 &\citep{Lasker2021-lr}\\
WISE W1   &    3.350 &  0.140 &  7.000e-03 &\citep{Lasker2021-lr}\\
Spitzer IRAC3.6   &    3.600 &  0.141 &  3.759e-04 & SEIP\\
Spitzer IRAC4.5   &    4.500 &  0.114 &  4.042e-04 & SEIP\\
WISE W2   &    4.600 &  0.116 &  0.011 & \citep{Lasker2021-lr}\\
Spitzer IRAC5.8   &    5.800 &  0.151 &  1.487e-03 & SEIP\\
Spitzer IRAC8.0   &    8.000 &  0.280 &  2.782e-03 & SEIP\\
Spitzer MIPS24   &   24.00 &  1.484 &  0.215 & SEIP\\
HERSCHEL PACS 70  &   70 &  6.768 &  3.606 & HPPSC\\
HERSCHEL PACS 160 &  160 &  20.976 &  7.838 & HPPSC\\
HERSCHEL SPIRE 250 &  250 &  37.6 &  7.900 & HPPSC\\
ALMA & 484 & 52.9 & 5.7 & Our work \\
ALMA & 869 & 20.5 & 2.1 & Our work \\
ALMA & 1110 & 9.0 & 0.9 & \add{\citep{Lu2018-fd}} \\
IRAM interferometer & 1350 & 5.6 & 1.1 & \add{\citep{Guilloteau1997-xe}}\\
ALMA & 2012 & 1.17 & 0.12 & \add{\citep{Lu2018-fd}} \\
ALMA & 2080 & 1.00 & 0.10 & Our work \\
VLA & 61685 & 0.076 & 0.011 & \add{\citep{Carilli1999-mo}} \\
VLA & 203940 & 0.220 & 0.043 & \add{\citep{Carilli1999-mo}} \\

\hline\hline
\end{tabular}
\caption{The data points used in this paper. From the left column, the name of the instrument, central wavelength in microns, measured flux, flux uncertainty, and sources. SEIP denotes Spitzer Enhanced Imaging Products (SEIP) Source List (doi: \href{https://www.ipac.caltech.edu/doi/irsa/10.26131/IRSA3}{10.26131/IRSA3}) and HPPSC denotes the Herschel/PACS Point Source Catalogue \citep{Marton2017-nl}. For ALMA photometric points, the absolute flux uncertainty (10\% of the flux) is added in the quadrature with the statistical uncertainty.}
\label{tab:taba1}
\end{table*}

\bsp	
\label{lastpage}
\end{document}